\documentclass[twocolumn]{aastex63}
\usepackage{color}
\usepackage{xspace}
\usepackage{glossaries}
\usepackage[single=true]{acro}
\usepackage{bm}        
\usepackage{amssymb}   
\usepackage{upgreek}
\usepackage{multirow}
\usepackage[normalem]{ulem}

\newcommand{\nus}{{\it NuSTAR}\xspace}
\newcommand{\suzaku}{{\it Suzaku}\xspace}
\newcommand{\fermi}{{\it Fermi}\xspace}
\newcommand{\chandra}{{\it Chandra}\xspace}
\newcommand{\comp}{{COMPTEL}\xspace}
\newcommand{\integralsat}{{\it INTEGRAL}\xspace}

\newcommand{\ls}{LS~5039\xspace}
\newcommand{\lsI}{LS~I+61$^\circ$303\xspace}
\newcommand{\psrb}{PSR~B1259-63/LS2883\xspace}

\newglossaryentry{grbs}{name={gamma-ray binary system},user1={Gamma-Ray Binary System},description={binary system with SED peaing in the gamma-ray energy band}}
\newglossaryentry{x}{name={X-ray},description={emission in the X-ray energy band}}
\newglossaryentry{g}{name={gamma ray},user1={gamma-ray},description={emission in the gamma-ray energy band}}
\newglossaryentry{lc}{name={light curve},description={general physical term}}
\newglossaryentry{lf}{name={Lorentz factor},description={general physical term}}
\newglossaryentry{suz}{name={Suzaku},description={satellite}}
\newglossaryentry{alf}{name={Alfv\'en},description={Hannes Alfv\'en}}

\received{}
\revised{}
\accepted{}
\submitjournal{ApJ}

\shorttitle{Spectral and Temporal Features using \nus and \fermi}
\shortauthors{Yoneda et al.}

\begin{document}

\title{Broad-Band High-Energy Emission of the Gamma-ray Binary System LS 5039:\\
Spectral and Temporal Features using \nus and \fermi Observations}

\correspondingauthor{Hiroki Yoneda}
\email{hiroki.yoneda@riken.jp}

\author[0000-0002-5345-5485]{Hiroki Yoneda}
\affiliation{RIKEN Nishina Center, 2-1 Hirosawa, Wako, Saitama 351-0198, Japan}
\affiliation{Department of Physics, The University of Tokyo, 7-3-1 Hongo, Bunkyo, Tokyo 113-0033, Japan}
\affiliation{Kavli Institute for the Physics and Mathematics of the Universe (WPI), The University of Tokyo Institutes for Advanced Study, The University of Tokyo, Kashiwa, Chiba 277-8583, Japan}

\author[0000-0002-7576-7869]{Dmitry Khangulyan}
\affiliation{Department of Physics, Rikkyo University, 3-34-1 Nishi Ikebukuro, Toshima, Tokyo 171-8501, Japan}

\author[0000-0003-1244-3100]{Teruaki Enoto}
\affiliation{Extreme natural phenomena RIKEN Hakubi Research Team, Cluster for Pioneering Research, RIKEN, Hirosawa 2-1, Wako, Saitama, 351-0198, Japan}

\author{Kazuo Makishima}
\affiliation{Kavli Institute for the Physics and Mathematics of the Universe (WPI), The University of Tokyo Institutes for Advanced Study, The University of Tokyo, Kashiwa, Chiba 277-8583, Japan}

\author{Kairi Mine}
\affiliation{Department of Physics, The University of Tokyo, 7-3-1 Hongo, Bunkyo, Tokyo 113-0033, Japan}

\affiliation{Kavli Institute for the Physics and Mathematics of the Universe (WPI), The University of Tokyo Institutes for Advanced Study, The University of Tokyo, Kashiwa, Chiba 277-8583, Japan}

\author[0000-0001-7263-0296]{Tsunefumi Mizuno}
\affiliation{Hiroshima Astrophysical Science Center, Hiroshima University, 1-3-1 Kagamiyama, Higashi-Hiroshima, Hiroshima, 739-8526, Japan}

\author{Tadayuki Takahashi}
\affiliation{Kavli Institute for the Physics and Mathematics of the Universe (WPI), The University of Tokyo Institutes for Advanced Study, The University of Tokyo, Kashiwa, Chiba 277-8583, Japan}
\affiliation{Department of Physics, The University of Tokyo, 7-3-1 Hongo, Bunkyo, Tokyo 113-0033, Japan}

\begin{abstract}
We report on detailed analysis of the hard X-ray and GeV gamma-ray spectra of \ls, 
one of the brightest gamma-ray binary system in the Galaxy.
The \nus observation covering its entire orbit in 2016 allowed us for the first time to study the orbital variability of the spectrum above 10 keV.
The hard X-ray spectrum is well described with a single power-law component up to 78 keV.
The X-ray flux showed a slight deviation from those observed previously with \suzaku in 2007.
The fast X-ray brightening observed with \suzaku, 
around the inferior conjunction,
was not observed in this observation.
We also analyzed 11 years of \fermi Large Area Telescope data of \ls.
The GeV spectrum around the inferior conjunction was well described with two non-thermal components;
a power law with a photon index of $\sim 3$
and a cut-off power law with a cutoff energy of $\sim 2$ GeV.
The orbital flux variability also changed gradually around a few GeV.
These results indicate that there are two emission components in the GeV band, and the dominant component above $\sim 1$ GeV does not depend on the orbital phase.
By combining these results, we update the spectral energy distribution of LS 5039 with the highest available statistics.
Theoretical models proposed so far cannot explain the obtained multi-wavelength spectrum, especially the emission from $\sim$ 1 MeV to $\sim$ 400 MeV,
and we discuss a possibility that particle acceleration in \ls is different from the shock acceleration.
\end{abstract}


\section{Introduction}
\label{sec_intro}
Gamma-ray binary systems are a subclass of compact binary systems established in the middle of 2000s  thanks to observations with ground based Cherenkov telescopes \citep{hessls50392005,PSRB1259HESS2005,2006Sci...312.1771A,2008ApJ...679.1427A}.
So far a handful of gamma-ray binary systems have been detected owing to
the recent development of GeV and TeV gamma-ray observations \citep{Dubus2013}.
These systems are known to consist of OB star primaries and compact-star secondaries.
However, compared to the thermal emissions of typical high-mass X-ray binaries (e.g. \citealt{Zdziarski2004,Becker_2007}),
their radiation spectra are completely different;
they show hard power-law spectra in the X-ray band, and their spectral energy distributions (SEDs) peak beyond 1 MeV.
The SEDs, dominated by non-thermal gamma-ray emission, indicate that efficient particle acceleration takes place in the gamma-ray binary systems.

Among a handful of gamma-ray binary systems known so far,
\ls is one of the brightest systems in the Galaxy.
It consists of an O-type primary star with a mass of $23 M_\odot$ \citep{casarespossible2005}
and a compact secondary, 
but whether the latter is a neutron star or a black hole is still unknown.
Its non-thermal emission has  a high bolometric luminosity 
of $\sim 1 \times 10^{36}$ erg s$^{-1}$, and the spectrum extends beyond TeV energies \citep{hessls50392005, collmarls2014}.
Based on a combined analysis of X-ray and TeV spectra,
it is suggested that particles are accelerated in this source
with an exceptionally high efficiency, close to, or even higher than, the maximum rate allowed in ideal magnetohydrodynamic sources \citep{khangulyan2008,takahashistudy2009}.

\ls has been detected from the radio to the TeV gamma-ray bands \citep{Paredes2000,casarespossible2005,takahashistudy2009,collmarls2014,fermi_lat_collaboration_2009,hessls50392005}, and
its non-thermal emission shows several distinct features.
\citet{collmarls2014} re-analyzed the \comp data, and
found a bright MeV gamma-ray source which is spatially consistent with \ls.
The source has been identified with \ls,
because it shows periodicity at the orbital period of \ls.
This MeV gamma-ray emission component dominates over those in the other energy bands.
Moreover, the GeV and TeV emission from \ls seems to be periodic over the system orbital period \citep{hess2006,fermi_lat_collaboration_2009}.
In the soft X-ray band, \citet{kishishitalongterm2009} revealed that the orbital-period-folded light curve has a good reproducibility on time scales of years.

Though several scenarios have been proposed so far,
it is suggested that \ls harbors a non-accreting neutron star rather than an accreting black hole
by several arguments e.g. the long term stability in soft X-rays \citep{kishishitalongterm2009} and 
spectral similarity to gamma-ray pulsars in the GeV band \citep{fermi_lat_collaboration_2009}.
A popular scenario at this moment is
that the compact object is a rotation-powered pulsar with strong pulsar winds \citep{Dubus2006b,takata2014,dubus2015}.
In this scenario, the pulsar is surrounded by a dense environment since the massive star emits strong stellar winds and ultraviolet photons.
Then, relativistic electrons in the pulsar winds interact with the surrounding media.
As a result, shocks are formed; particles from the pulsar wind are accelerated and produce non-thermal emission via the synchrotron and inverse Compton channels \citep{1997ApJ...477..439T,1999APh....10...31K,2007MNRAS.380..320K}.

This scenario can explain the broad-band spectrum in the X-rays and GeV/TeV bands \citep[see, e.g.,][]{takata2014,dubus2015}.
However, 
the bright MeV emission of \ls has not yet been explained successfully either by this popular scenario, or by any alternative ideas currently elaborated.
Similar bright MeV emission was also detected or suggested from other gamma-ray binary systems e.g. \lsI \citep{Dijk1996,Tavani1996} and during GeV gamma-ray flares from \psrb \citep{2011ApJ...736L..11A,2015ApJ...811...68C,2018ApJ...863...27J}.
Surprisingly, the apparent luminosity of the flares from PSR B1259-63 is very close to or even higher than the limit imposed by the pulsar spin-down luminosity \citep{2020MNRAS.497..648C}.
Again, currently no scenario explains these MeV/sub-GeV components successfully.
These examples show that the interpretation of MeV/sub-GeV emission in gamma-ray binary systems is a challenging subject in several aspects.

Recently, an alternative idea has been proposed by \citet{yoneda2020}; \ls contains a magnetar rather than a rotation-powered pulsar.
By performing timing analysis using hard X-ray data of \suzaku and \nus,
they found a sign of 9 sec pulsation from \ls.
The obtained spin period candidate and its derivative yield the spindown luminosity two orders of magnitude lower than the bolometric luminosity of \ls.
Then, they proposed that the magnetic energy of the neutron star in \ls drives its non-thermal activity,
which requires magnetar-class strong magnetic fields to explain its luminosity.
\cite{Volkov2021} also performed a detailed timing/spectral analysis, by mainly using the same \nus data.
The magnetar binary hypothesis was also proposed in \lsI, based on detection of a X-ray flare similar to those observed in isolated mangetars \citep{torresmagnetarlike2012}.

In this work,
we challenge the mystery of the non-thermal emission of \ls,
through a study of its spectral properties in the GeV and hard X-ray bands,
which are adjacent in energy to the MeV band.
Although a deep X-ray observation was performed with the Hard X-ray Detector (HXD) onboard \suzaku,
and valuable results have been obtained \citep{takahashistudy2009,kishishitalongterm2009,yoneda2020},
the HXD background was not yet low enough to derive high-quality hard X-ray spectra of \ls.
In order to investigate the spectrum above 10 keV accurately,
we analyze the \nus data of \ls with an exposure time of $\sim 350$ ks in Section~\ref{sec_nustar}.

In Section~\ref{sec_fermi}, we analyze 11 years of \fermi data of \ls.
Previous works have reported several spectral features e.g. a possibility that the GeV spectrum around the inferior conjunction has a hump around 2 GeV, or the orbital-phase dependence is different in the 0.2--3 GeV and  3--20 GeV bands \citep{fermi_lat_collaboration_2009, Hadasch2012, Chang2016}.
We investigate these features in more detail using the latest data.
Finally, by combining the results from the \nus and \fermi data,
we update the SED of \ls with the highest available statistics.
We compare it with previous spectral models and clarify problems in our current understanding in Section~\ref{sec_discussion}.
Finally, we conclude this work in Section~\ref{sec_conclusion}.

\section{Spectral Analysis in Hard X-ray band}
\label{sec_nustar}

\subsection{{\it NuSTAR} Observation and Data Reduction}
\ls was observed with \nus (OBSID: 30201034002) from 2016 September 1 to 5
for a gross exposure of 346 ks,
which is roughly a full orbital period of \ls.
This data set was already utilized by \citet{yoneda2020}, where the hint of 9 sec pulsation was obtained.
The data reduction and analysis were performed 
by the {\tt NuSTARDAS}
version 1.8.0, and {\tt NuSTAR CALDB} version 20180312.
We extracted source events
from a circular area centered at the source position with a radius of 30 arcsec.
The number of 3--78 keV source events obtained from the two focal plane modules, FPMA and FPMB,
are 18656 and 18212, respectively.

\begin{figure}[!htbp]
\begin{center}
\includegraphics[width = 8.0 cm]{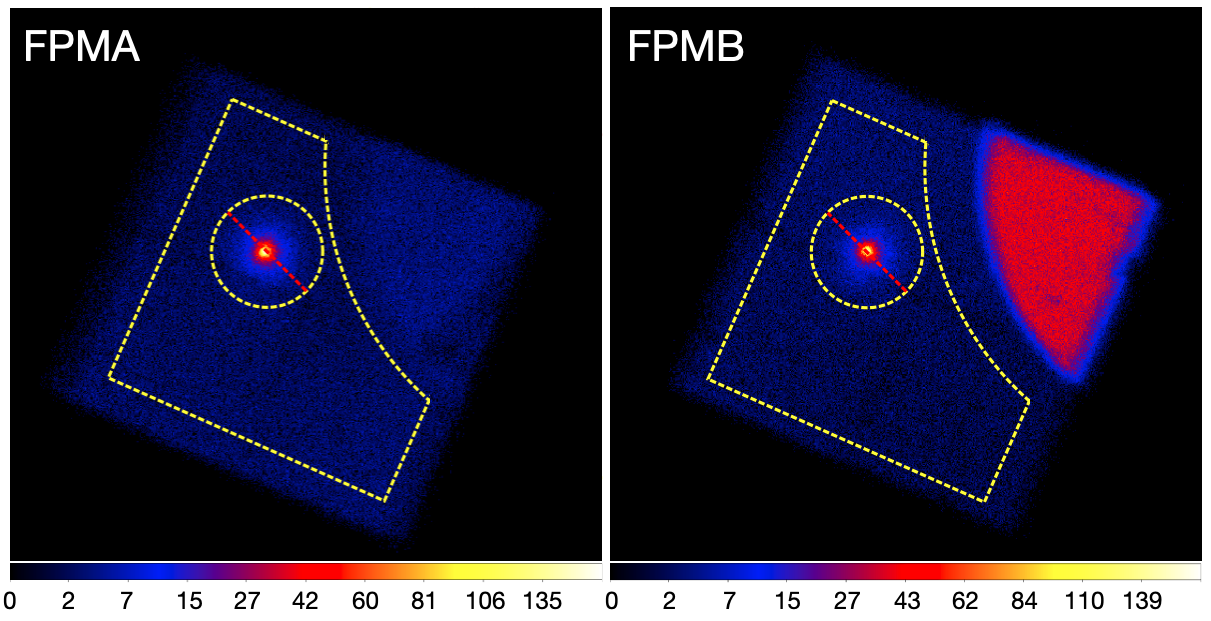}
\end{center}
\caption{3--78 keV count maps obtained from FPMA (left) and FPMB (right).
The yellow dashed line represents the background region selected in the spectral analysis of \nus.
Note that the circle region with the red dashed line is not in the background region.
The upper right region with high photon counts was produced by the stray light from the nearby bright X-ray source GX 17+2.
This region is not used for the background estimation.
The numbers of events extracted from the background region are 60323 for FPMA and 53865 for FPMB.
}
\label{fig_back_nustar}
\end{figure}

A solar coronal mass ejection (CME) was observed during this observation.
Thus we optimized the calculation of the South Atlantic Anomaly (SAA) passages
because the CME increases the background level close to the SAA.
We choose a parameter set ({\tt saacalc=2 saamode=OPTIMIZED tentacle=yes}) that removes high background intervals as much as possible without sacrificing the net exposure time (165 ks) significantly.
The background region was defined as shown in Figure~\ref{fig_back_nustar}.
The contribution of the background to the spectrum is estimated to be about 4 \%.

\subsection{Spectral Analysis in 3 -- 70 keV}
\label{sec_nustar_ana_allene}
In the spectral analysis,
the absorption by the interstellar medium and by the material in the observed system
should be considered, especially in the low energy band.
The equivalent hydrogen column density would be determined by a low energy X-ray spectrum ($<$ 1 keV),
but the \nus data do not cover the X-ray band below 3 keV.
We hence utilized the column density measured by \citet{kishishitalongterm2009}, 
and we fixed it to $N_\mathrm{H} = 0.77 \times 10^{21}~\mathrm{cm^{-2}}$ using the {\tt phabs} model in XSPEC.
Although \citet{kishishitalongterm2009} also reported that $N_\mathrm{H}$ varied from $0.64 \times 10^{21}~\mathrm{cm^{-2}}$ to $0.79 \times 10^{21}~\mathrm{cm^{-2}}$ depending on observation epochs,
we confirmed that this slight variability yields negligible systematic errors of parameters estimated in the following spectral analysis, compared to their statistical errors.

\begin{figure}[!htbp]
\begin{center}
\includegraphics[width = 8.5 cm]{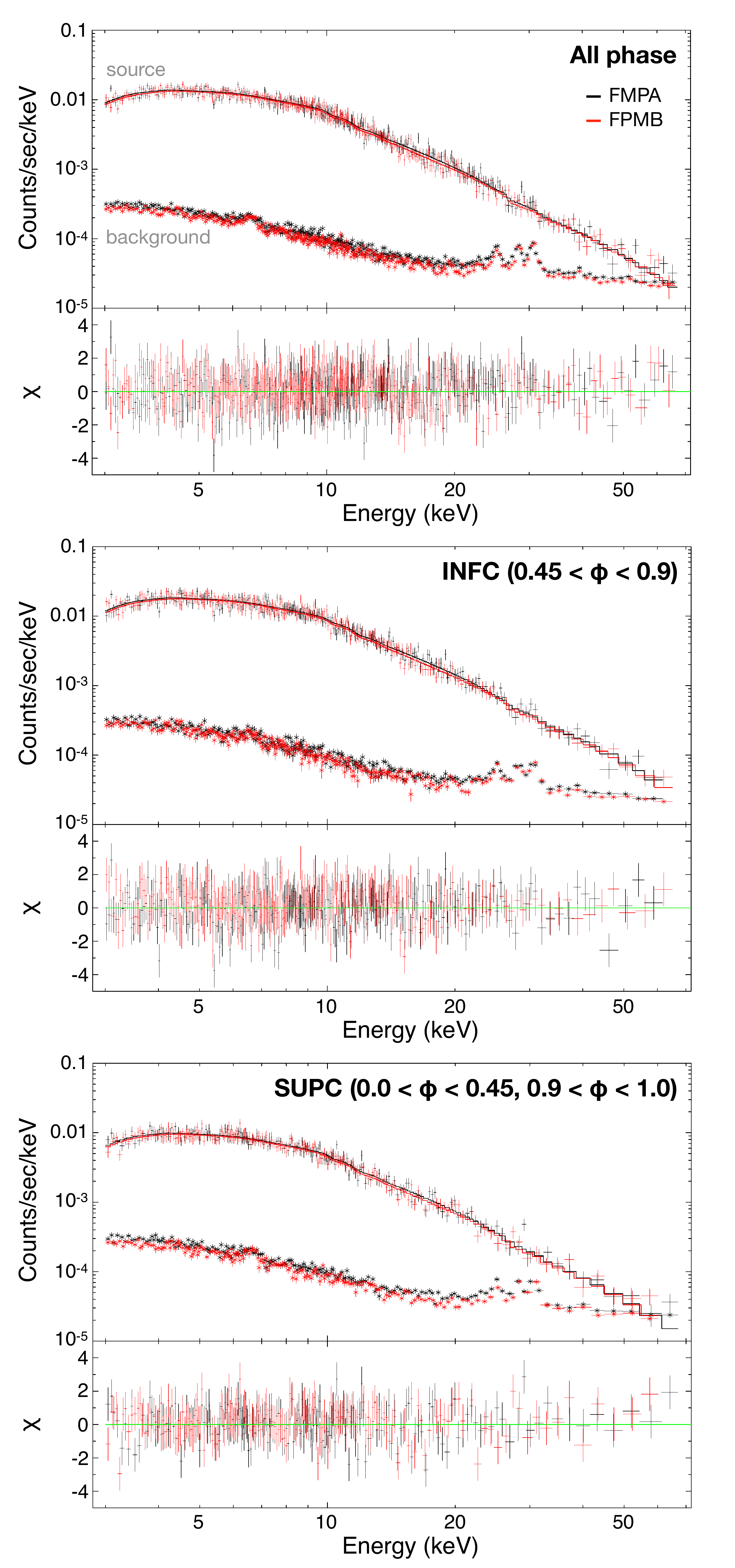}
\end{center}
\caption{The time-averaged \nus spectrum (top) and the spectra in INFC (middle) and SUPC (bottom). The black/red lines indicate FPMA/FPMB, respectively. The background spectra are shown below the source spectra.}
\label{fig_nus_wide_band_fit}
\end{figure}

\begin{figure}[!htbp]
\includegraphics[width = 8.0 cm]{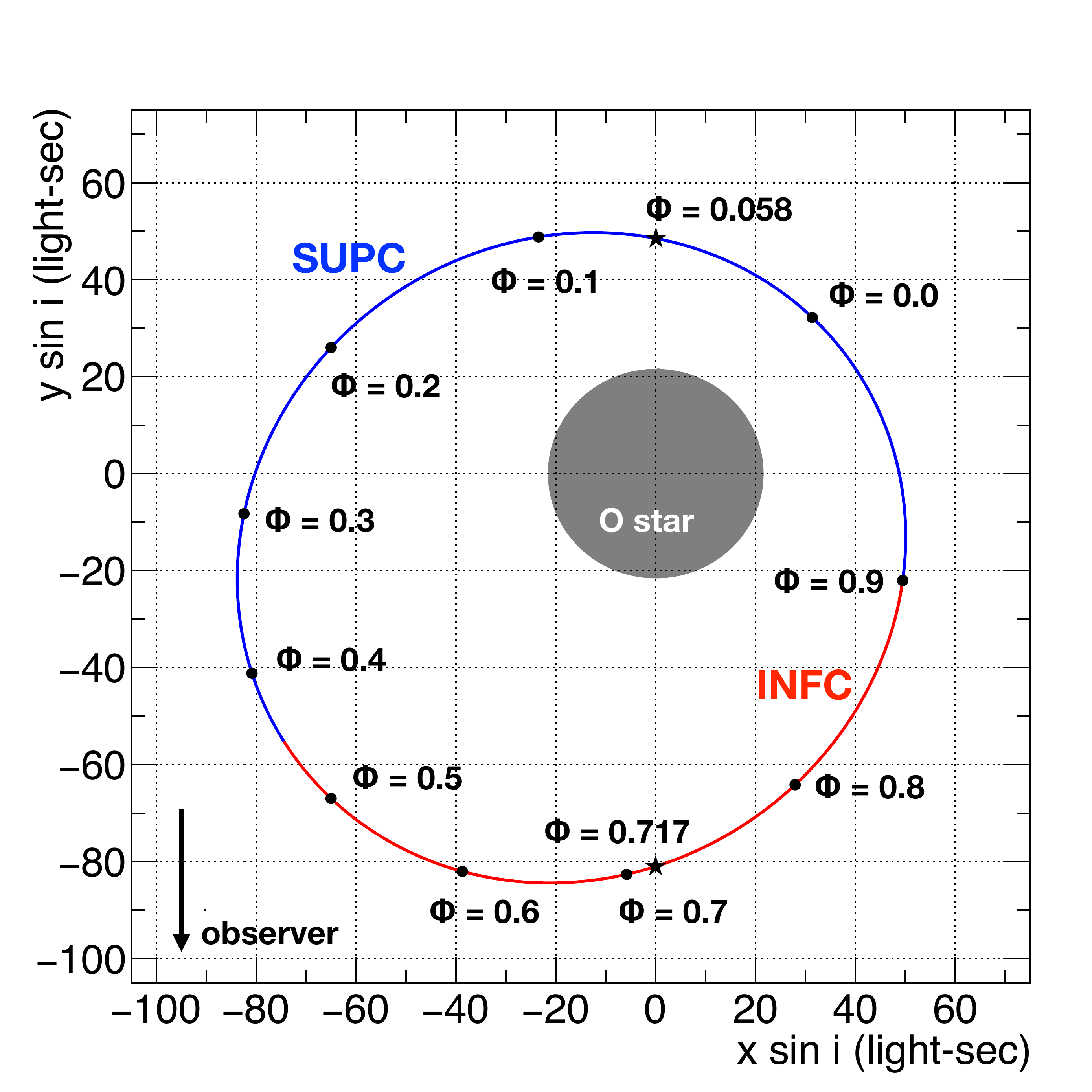}
\caption{Orbital geometry of \ls, after \cite{casarespossible2005}, projected on the orbital plane.
Here, $\phi$ is the orbital phase, and $i$ is the inclination angle.
$\phi = 0$ corresponds to the periastron of \ls.
The compact star is the closest to the observer at $\phi = 0.670$ (inferior conjunction) and the farthest at $\phi = 0.046$  (superior conjunction).
Here, the mass of the compact star is assumed to be 1.4 $M_\odot$.
The mass and radius of the companion star are set to 2.9 $M_\odot$ and 9.3 $R_\odot$, respectively \citep{casarespossible2005}.
}
\label{fig_ls_orbit}
\end{figure}

Figure~\ref{fig_nus_wide_band_fit} top shows the spectrum of \ls using all events of the \nus data.
It is described very well with a single power-law model with a photon index of $1.63\pm{0.01}$.
The best-fitting parameters are shown in Table~\ref{tab_spectrum_3_70_keV_nustar}.
In order to investigate the spectral change depending on the orbital phase,
we defined the orbital phase around inferior conjunction (INFC) as $0.45 < \phi < 0.9$, 
and superior conjunction (SUPC) as $0.0 < \phi < 0.45$ and $0.9 < \phi < 1.0$,
after \citet{hess2006}.
These intervals are indicated in Figure~\ref{fig_ls_orbit}.
To compare our hard X-ray analysis with the previous results,
the orbital phase was calculated using the orbital parameters by \citet{casarespossible2005}.
The spectra in the INFC and SUPC phases are presented in the lower two panels in Figure~\ref{fig_nus_wide_band_fit}.
Thus, both spectra are well described by a single power-law model.
Here we rebinned the spectra so that each bin contains more than 30 counts, and used chi-square statistics for the spectral fitting.

\begin{table}[!htb]
\caption{The best-fitting parameters obtained from 3--70 keV \nus data.}
\label{tab_spectrum_3_70_keV_nustar}
  \begin{center}
    \begin{tabular}{c c c c} \hline
      & Photon index & Flux (3--70 keV)$^\dagger$ & $\chi^2/$dof\\ \hline
      INFC & $1.61\pm0.01$ & $3.19\pm{0.04}$ & $514.6/526$\\
      SUPC & $1.64\pm0.02$ & $1.65\pm{0.03}$ & $334.5/375$\\
      All & $1.63\pm0.01$ & $2.34\pm{0.02}$ & $676.0/677$\\ \hline
    \end{tabular}
\end{center} 
{\small $\ast$: Errors correspond to 1 $\sigma$ confidence interval. \\
$\dagger$: Its unit is $10^{-11}~\rm erg~cm^{-2}~s^{-1}$.}
\end{table}

\subsection{Comparison with the \suzaku/XIS Results}
\label{sec_comp_nustar_suzaku}
In order to investigate the dependence of spectral parameters on the orbital phase,
we divided the data into subsets corresponding to different orbital phase intervals, and analyzed the spectrum of each set separately.
In addition to this, we compare our result with the previous \suzaku observation \citep{takahashistudy2009,kishishitalongterm2009}.
Considering the energy range of the \suzaku XIS,
we focus on the spectrum below 10 keV in this subsection.

\begin{figure}[!htbp]
\begin{center}
\includegraphics[width = 8.5 cm]{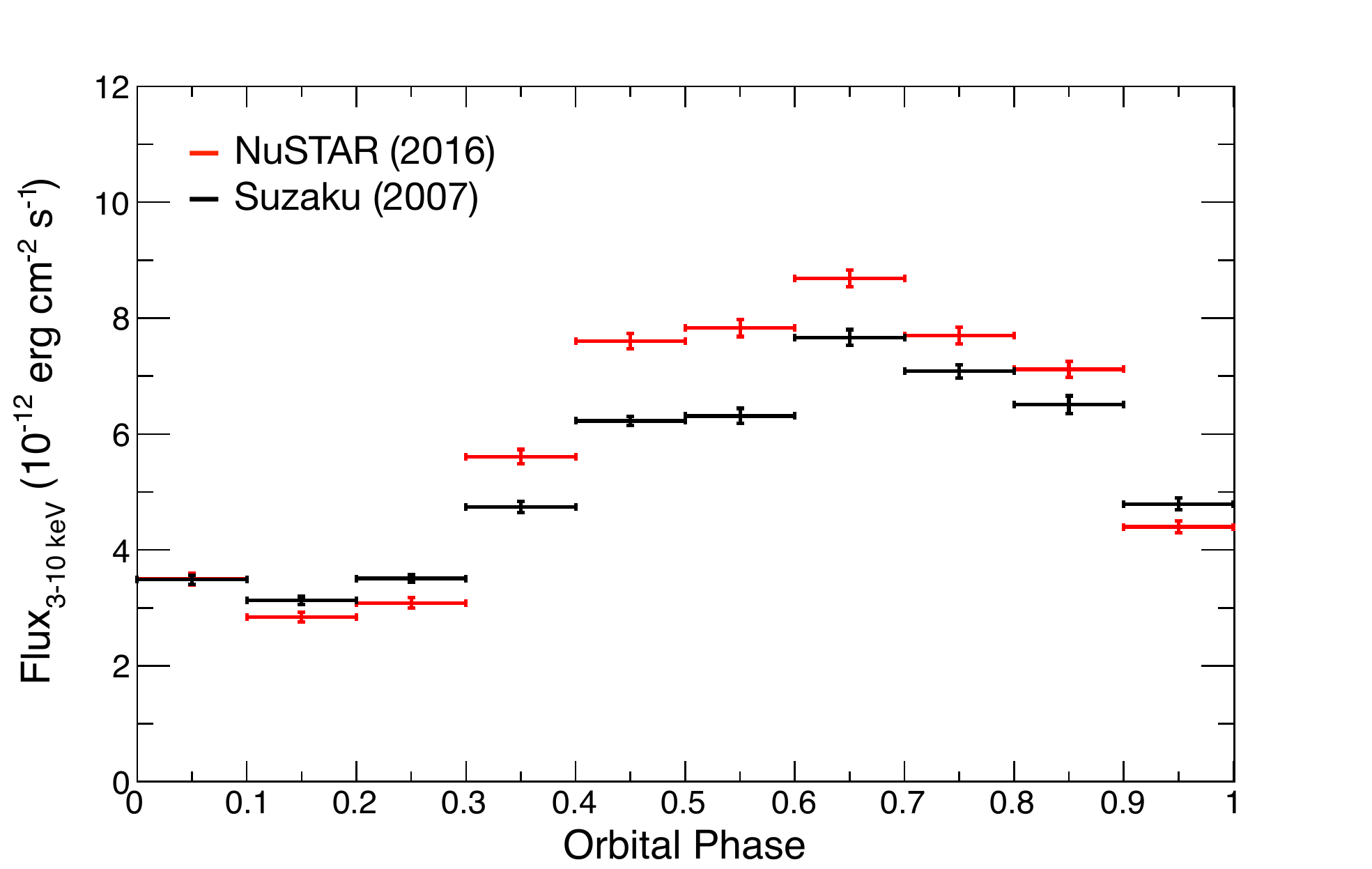}
\includegraphics[width = 8.5 cm]{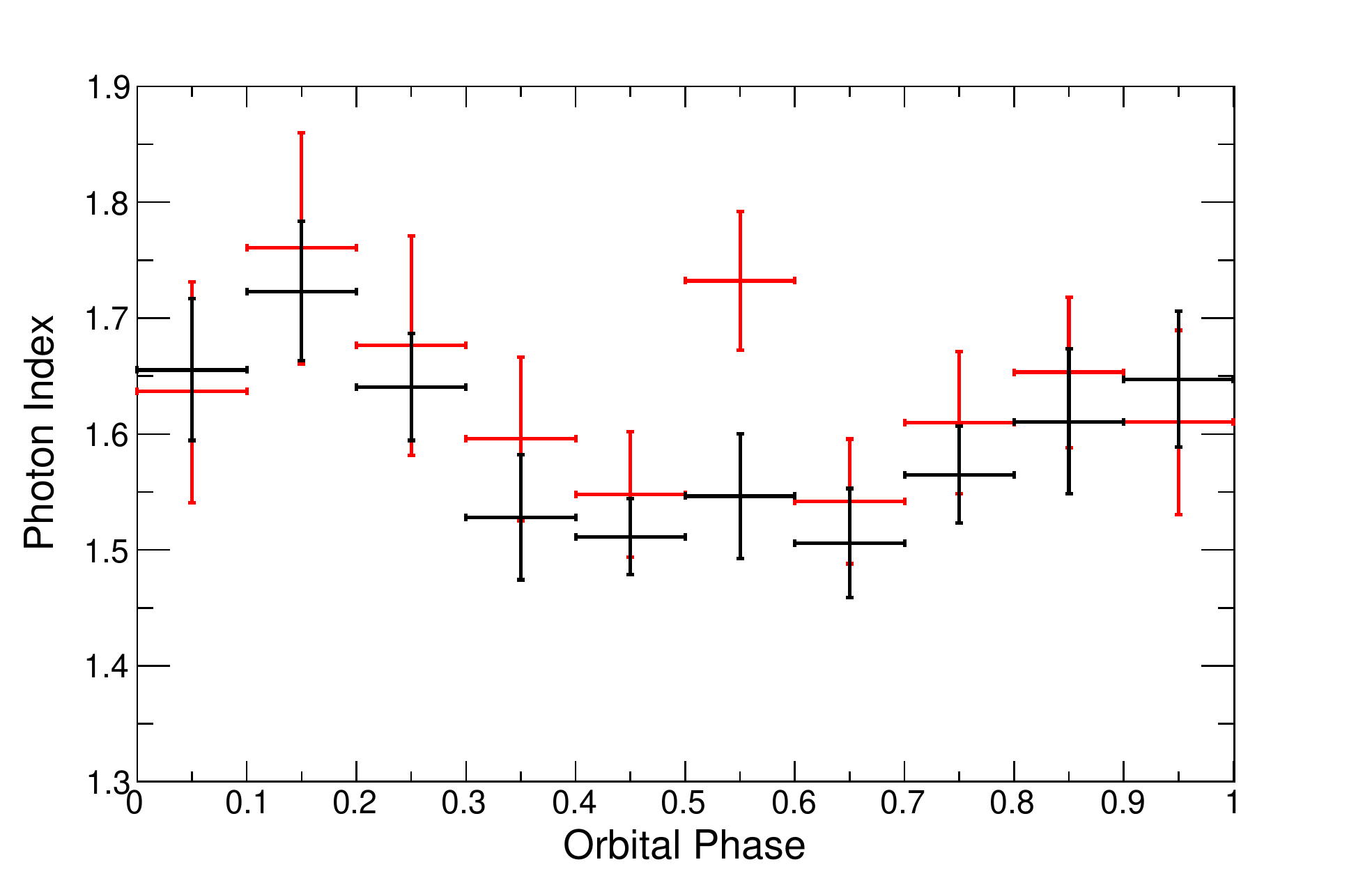}
\end{center}
\caption{Orbital variations of the flux (top) and the photon index (bottom).
The 3--10 keV \nus results in red are compared with those from the 3--10 keV \suzaku data in black.}
\label{fig_flux_1_10_keV}
\end{figure}

For the above purpose,
we first considered ten orbital-phase intervals of \(\sim34\rm\,ks\) each.
The results obtained from these intervals are described in Table~\ref{tab_spectrum_1_10_keV} and shown in Figure~\ref{fig_flux_1_10_keV}.
For the comparison with the \nus observation,
we also analyzed 3--10 keV \suzaku/XIS data in the same way as \citet{takahashistudy2009} and measured 3--10 keV flux in different orbital phase intervals.
Then, the flux becomes minimum and maximum at $\phi$ = 0.1--0.2 and $\phi$ = 0.6--0.7,
respectively, in agreement with the \suzaku light curve.
While the flux at the  minimum is consistent with that measured with \suzaku, the maximum \nus flux 
is higher by 10--20 \% than that with \suzaku.
The difference is larger than the uncertainty due to the cross calibration between \suzaku and \nus \citep[10\% at most,][]{cross_calibaration}.
As a result,
the flux difference between \nus and \suzaku is significant around INFC ($0.45 < \phi < 0.9$).

We also investigated changes of the photon index.
The bottom panel of Figure~\ref{fig_flux_1_10_keV} shows the photon indices at different orbital phases.
They depend on the orbital phases in both observations. 
The indices measured with \nus are mostly higher by $\sim 0.04$ than those with \suzaku,
but they are consistent within their statistical errors except for $\phi =$ 0.5--0.6.
Though the photon indices at INFC/SUPC with \nus are consistent with each other within the statistical errors (see Table~\ref{tab_spectrum_3_70_keV_nustar}),
it is possibly due to the large photon index at $\phi =$ 0.5--0.6.
These results are robust to changes in $N_\mathrm{H}$ within its uncertainty.
Note that the \suzaku observation also provides the hard X-ray data above 10 keV,
but it is difficult to investigate the orbit-by-orbit variability above 10 keV due to large statistical errors (see Appendix~\ref{ap_hxd}).

\begin{table*}[!htbp]
  \begin{center}
\caption{Results of spectral analysis in 3--10 keV.}
\label{tab_spectrum_1_10_keV}
    \begin{tabular}{c  c c c | c c c} \hline
        & \multicolumn{3}{c|}{\nus (3--10 keV)}& \multicolumn{3}{c}{\suzaku (3--10 keV)} \\ \hline
      Orbital Phase & Photon index & Flux (3--10 keV)$^\dagger$ & $\chi^2/$dof & Photon index & Flux (3--10 keV)$^\dagger$ & $\chi^2/$dof \\ \hline
      INFC & $1.61\pm{0.03}$ & $7.92 \pm 0.07$ & $311.1/317$ & $ 1.53 \pm 0.02 $ & $ 6.86 \pm 0.06 $ & $936.8/989$\\ 
      SUPC & $1.62\pm{0.03}$ & $4.26 \pm 0.04$ & $169.5/227$ & $ 1.62 \pm 0.02 $ & $ 4.07 \pm 0.03 $ & $1172.1/1141$\\ 
      All Phase & $1.62\pm0.02$ & $5.94 \pm 0.04$ & $341.5/346$ & $ 1.58 \pm 0.02 $ & $ 5.06 \pm 0.03 $ & $1913.7/1874$\\ \hline     
      0.0--0.1&  $ 1.64_{-0.10}^{+0.09} $  & $ 3.50 \pm 0.10 $ & $ 36.2/40 $  & $ 1.66 \pm 0.06 $ & $ 3.49 \pm 0.08 $ & $ 165.4/182 $\\
      0.1--0.2&  $ 1.76 \pm 0.10 $         & $ 2.84 \pm 0.09 $ & $ 34.2/35 $  & $ 1.72 \pm 0.06 $ & $ 3.13 \pm 0.07 $ & $ 181.6/197 $\\
      0.2--0.3&  $ 1.68 \pm 0.09 $         & $ 3.09 \pm 0.09 $ & $ 40.4/37 $  & $ 1.64 \pm 0.05 $ & $ 3.51 \pm 0.06 $ & $ 322.7/297 $\\
      0.3--0.4&  $ 1.60 \pm 0.07 $         & $ 5.61 \pm 0.12 $ & $ 73.0/67 $  & $ 1.53 \pm 0.05 $ & $ 4.74 \pm 0.10 $ & $ 162.6/191 $\\
      0.4--0.5&  $ 1.55 \pm 0.05 $         & $ 7.60 \pm 0.13 $ & $ 104.9/98 $ & $ 1.51 \pm 0.03 $ & $ 6.23 \pm 0.08 $ & $ 488.5/469 $\\
      0.5--0.6&  $ 1.73 \pm 0.06 $         & $ 7.83 \pm 0.15 $ & $ 68.2/86 $  & $ 1.55 \pm 0.05 $ & $ 6.31 \pm 0.13 $ & $ 210.7/175 $\\
      0.6--0.7&  $ 1.54 \pm 0.05 $         & $ 8.68 \pm 0.14 $ & $ 97.0/106 $ & $ 1.51 \pm 0.05 $ & $ 7.66 \pm 0.14 $ & $ 221.2/220 $\\
      0.7--0.8&  $ 1.61 \pm 0.06 $         & $ 7.70 \pm 0.14 $ & $ 93.1/91 $  & $ 1.56 \pm 0.04 $ & $ 7.08 \pm 0.11 $ & $ 294.8/282 $\\
      0.8--0.9&  $ 1.65_{-0.07}^{+0.06} $  & $ 7.12 \pm 0.14 $ & $ 60.4/77 $  & $ 1.61 \pm 0.06 $ & $ 6.51 \pm 0.15 $ & $ 133.1/131 $\\
      0.9--1.0&  $ 1.61 \pm 0.08 $         & $ 4.40 \pm 0.10 $ & $ 63.4/59 $  & $ 1.65 \pm 0.06 $ & $ 4.79 \pm 0.10 $ & $ 187.1/185 $\\ \hline
    \end{tabular}
\end{center}
{\small $\ast$: Errors represent 1 $\sigma$ confidence interval. $\dagger$: Their units are $10^{-12}~\mathrm{erg~cm^{-2}~s^{-1}}$. 
}
\end{table*}

\begin{figure}[!htbp]
\begin{center}
\includegraphics[width = 8.5 cm]{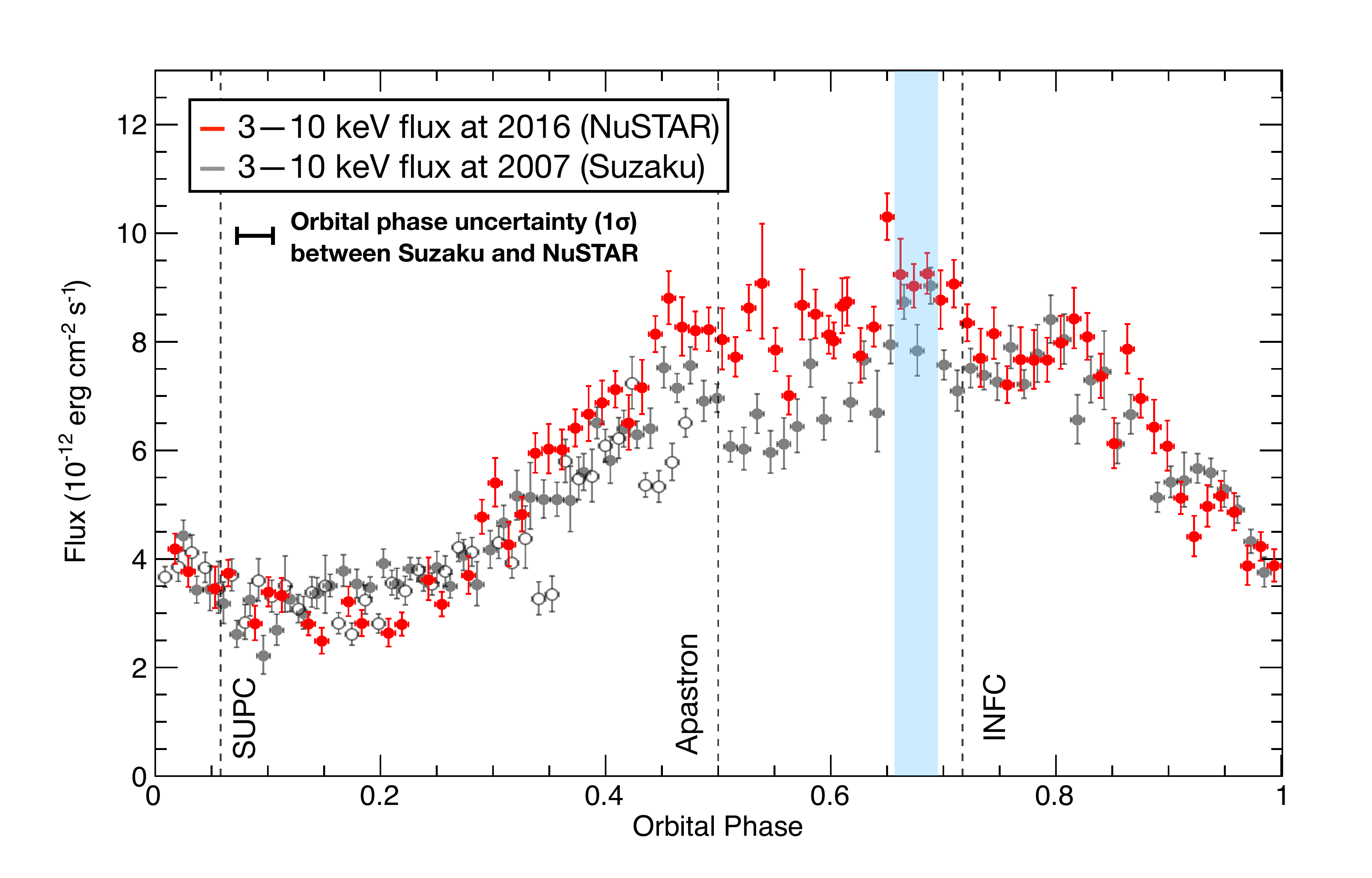}
\end{center}
\caption{Orbital light curve in 3--10 keV with a bin width of 4 ks.
The red and gray points indicate the results from \nus and \suzaku/XIS, respectively.
The gray open circles represents the second orbit of \ls observed in the \suzaku observation. 
The cyan region represents the orbital interval where the fast X-ray brightening was observed in \citet{kishishitalongterm2009}.
The horizontal bar represents the uncertainty of the orbital phase of the \nus data relative to \suzaku observation time, 
due to the orbital period uncertainty reported in \cite{casarespossible2005}.
}
\label{fig_lightcurve_hardX}
\end{figure}

In order to investigate short-time variability,
we also calculated the 3--10 keV flux in 4 ks time intervals,
by fitting the 3--10 keV spectrum in each interval with a single power-law model.
Here we used the C-statistic for the fitting because the number of photons in each interval is small.
Figure~\ref{fig_lightcurve_hardX} shows the resulting orbital light curve,
which is similar to Figure~\ref{fig_flux_1_10_keV},
but divided with the finer time steps.
The flux changes gradually, and no flares are observed.
Around $\phi = 0.8$,
the light curve shows a potential flux increase,
which was also observed by \citet{kishishitalongterm2009}. 
However, at $\phi = 0.70$ of the \nus observation,
Figure~\ref{fig_lightcurve_hardX} reveals no sharp spike 
that was observed by \citet{kishishitalongterm2009} in {\it Chandra} and the \suzaku data collected in 2004 and 2007, respectively.
This suggests that the spike structure in the light curve either varies from orbit to orbit or is limited to the low energy band ($<$ 3 keV) which {\it NuSTAR} does not cover.
To examine the latter possibility, we also analyzed the \suzaku/XIS data above 3 keV in the short time intervals. 
Then, as shown in Figure~\ref{fig_lightcurve_hardX}, the spike becomes less prominent even in the \suzaku data above 3 keV. Therefore, the latter of the above alternatives is favored.

Figure~\ref{fig_lightcurve_hardX} also shows the flux difference between the two observations in the orbital phase of 0.5--0.7, while the light curve is stable over the other phases.
Although \citet{kishishitalongterm2009} reported the reproducibility of the X-ray light curve in years,
most of the X-ray observations compared with \suzaku in their work lie in the orbital phase of 0.3--0.55.
Thus, we speculate that the X-ray emission from apastron to the inferior conjunction has orbit-by-orbit variability.
Additional X-ray observations around this orbital phase would be essential to confirm this new possibility.

\section{Spectral Analysis in GeV gamma-ray band}
\label{sec_fermi}

\subsection{\fermi Observation and Data Reduction}
The \fermi Large Area Telescope (LAT) data set used in this work 
spans from 2008 August 4 to 2019 September 21.
It was reduced and analyzed using the \fermi Science Tools 1.2.23\footnote{Details of the Science Tools are described in \url{https://github.com/fermi-lat/Fermitools-conda}.}.
We used Reprocess Pass 8 data classified as event class ``P8 source ({\tt evclass=128})'' and event type ``FRONT+BACK ({\tt evtype=3})'' \citep{pass8_1,pass8_2}.
In order to reduce the contamination from Earth limb emission,
we selected events which were detected at a zenith angle less than 90 degrees.
In the spectral analysis,
the {\tt P8R3\_SOURCE\_V2} instrument response functions were used.
For the modeling of the Galactic diffuse emission and isotropic backgrounds,
{\tt gll\_iem\_v07.fits} and {\tt iso\_P8R3\_SOURCE\_V2\_v1.txt}\footnote{These background models are available from the Fermi Science Support Center: \url{https://fermi.gsfc.nasa.gov/ssc/data/access/lat/BackgroundModels.html}.} were used.

A bright GeV pulsar, PSR J1826-1256, is located $\sim 2$ degrees from \ls.
In order to reduce the contamination from this source,
we followed previous studies \citep{fermi_lat_collaboration_2009,Hadasch2012},
and discarded the events whose arrival times are close to the peaks of the pulse emission of PSR J1826-1256.
For this purpose,
we analyzed the 11 years of \fermi LAT data for the $\sim 0.11$ sec pulsations from PSR J1826-1256, and obtained its frequency as well as 1st- and 2nd-order time derivatives.
We divided the data every two years and determined these parameters for each epoch (Appendix~\ref{sec_pulsar_elim} for details).
Since this pulsar has two sharp peaks in its pulse profile,
we excluded the events which fall on $0.455 < \phi_\mathrm{psr} < 0.605 ~\mathrm{or}~ 0.925 < \phi_\mathrm{psr} < 1.075$,
where $\phi_\mathrm{psr}$ is the pulse phase of PSR J1826-1256, defined in Appendix~\ref{sec_pulsar_elim}.
By discarding these events,
the number of the events was reduced by 30\%.
To account for this, we multiplied a scaling factor 0.7, to the normalization factors of all sources in the following binned likelihood analysis.

\subsection{Spectrum of \ls in the GeV gamma-ray band}
\label{sec_fermi_ana}
\begin{figure*}[!htbp]
    \begin{center}
    \includegraphics[width = 8.8 cm]{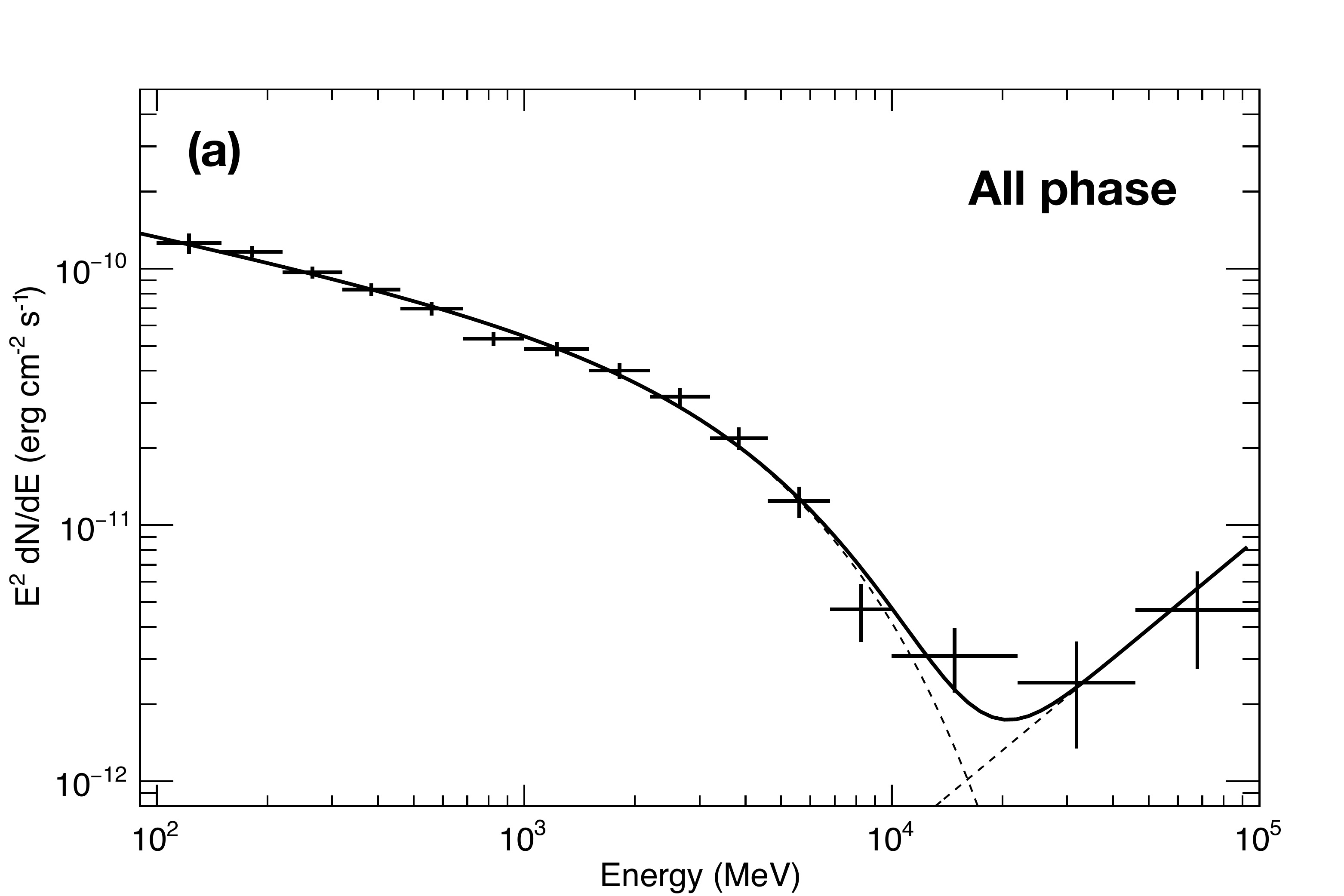}
    \includegraphics[width = 8.8 cm]{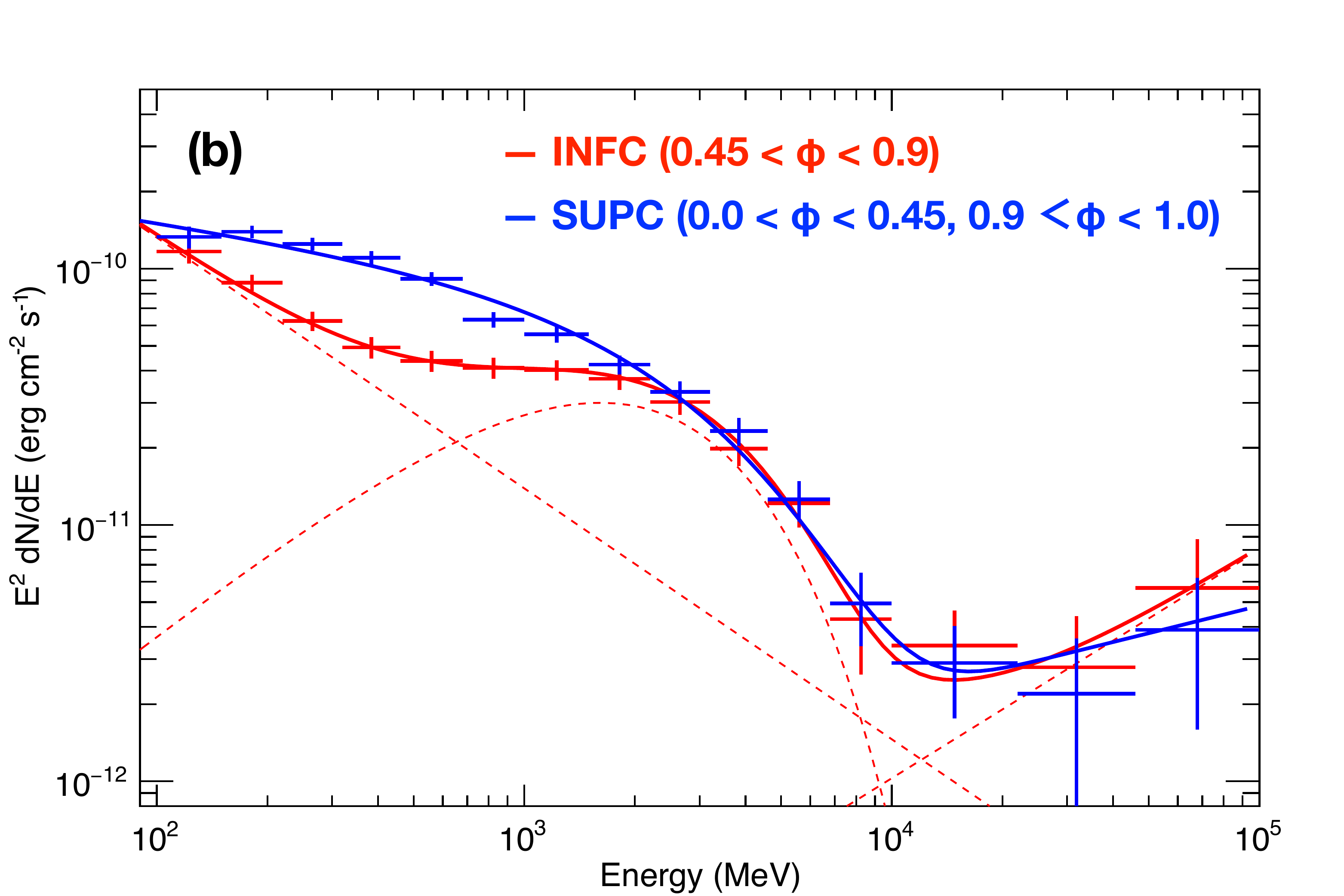}
    \end{center}
        
\caption{GeV spectra of \ls obtained from the \fermi pass 8 data. 
(a) The spectrum averaged over all the orbital phase.
(b) Those derived from the INFC (red) and SUPC (blue) intervals.
In the phase-averaged and SUPC spectra, 
the solid lines represent the best-fitting models of a sum of an exponential-cutoff power law and a single power law.
In the INFC spectrum, the best-fit of a sum of an exponential-cutoff power-law and two power-law components is shown as the red line.
}
\label{fig_fermi_fit}
\end{figure*}

Figure~\ref{fig_fermi_fit}a shows the orbit-averaged spectrum of LS 5039 over the entire 11 years. 
To model it, we employed a power law with an exponential cutoff which is defined as
\begin{eqnarray}
\label{eq_model_ls}
\frac{dN}{dE} = N_0 \left(\frac{E}{1~\mathrm{GeV}}\right)^{-\Gamma} \exp\left(-\frac{E}{E_\mathrm{cut}}\right)~,
\end{eqnarray}
where $\Gamma, E_\mathrm{cut}$ and $N_0$ are the photon index, the cutoff energy and the normalization factor, respectively.
The binned maximum likelihood method was used for the spectral fitting.
The bin width and the size of count maps were set to 0.125 degrees and 20$\times$20 deg$^2$ respectively.
The energy range and intervals were defined as 100 MeV -- 100 GeV and 12 bins per decade in logarithmic scale, respectively.
The energy dispersion correction was applied in the fitting.
We included all sources that are within 20 degrees of the Region of Interest (ROI) center and listed in the 4FGL catalog \citep{fermi4thcatalog}.
Their spectral models were selected to be the same as described in the catalog.
The spectral parameters of the sources which are within 5 degrees of the ROI center were allowed to vary freely.
In addition, we treated the parameters of PKS 1830-211 in the same way,
because it is known as a bright and variable blazar \citep{PKS2015} and apart from \ls by $\sim$ 6.5 degrees.
We also set free the normalization and the photon index of the Galactic diffuse model, as well as the normalization of the isotropic background model.
Significance of source detection was evaluated using $TS$ (Test Statistics) that is determined as $-2 \ln (L_0/L_1)$, where $L_{0/1}$ is the maximum likelihood without/with the source model \citep{Mattox1996}.

The systematic uncertainties of the fitting were estimated using the so-called bracketing Aeff method.
In this method,
we change the effective area by multiplying it by $1 \pm \epsilon$ below a certain energy $E_{\mathrm{switch}}$ and by $1 \mp \epsilon$ above.
Then, the maximum changes in the best-fit parameters, in response to $\epsilon$, are regarded as the systematic errors associated with them.
We set $\epsilon \simeq 0.03$ and $E_{\mathrm{switch}}$ as several values i.e. near $E_\mathrm{cut}$ in Eq.~\ref{eq_model_ls} and 20 GeV\footnote{Note that $\epsilon$ depends on the energy. See \url{https://fermi.gsfc.nasa.gov/ssc/data/analysis/scitools/Aeff_Systematics.html} for details.}.
Moreover, we also considered systematic errors due to the uncertainty of the Galactic diffuse model.
Following \cite{2PC2013},
we increase/reduce its normalization by 6\%,
and again considered the changes in the best-fit parameters as the systematic errors.
In Table~\ref{tab_fermi_fit_allphase} and \ref{tab_fermi_fit},
we described the maximum value in the obtained systematic errors.
In most cases, the error due to the Galactic diffuse model uncertainty is dominant.

As shown in Figure~\ref{fig_fermi_fit}a,
the orbit-averaged spectrum of \ls was described successfully by Eq.~\ref{eq_model_ls} below 10 GeV.
Above 10 GeV, the fit leaves a prominent excess, which is considered to be the low energy tail of the TeV component.
To model this excess, we added a power-law component defined as
\begin{eqnarray}
\label{eq_model_ls_TeV}
\frac{dN}{dE} = N_{\mathrm{TeV}} \left(\frac{E}{30~\mathrm{GeV}}\right)^{-\Gamma_{\mathrm{TeV}}}~,
\end{eqnarray}
and compared the results.
The best-fitting parameters using the 100 MeV--100 GeV gamma rays
are described in Table~\ref{tab_fermi_fit_allphase}.
By adding the power-law component, 
the $TS$ value was improved by $\Delta TS = 35$.
Its corresponding chance probability is $2.5 \times 10^{-8}$, because we added two free parameters and $\Delta TS$ obeys a chi-square distribution with 2 degrees of freedom under the null hypothesis.
Thus, the TeV component is significantly detected.
Finally, we obtained
$\Gamma \simeq 2.3$ and $E_\mathrm{cut} \simeq 4.8$ GeV for the component described by Eq~\eqref{eq_model_ls}.

\begin{table}[!htbp]
\caption{Best-fitting parameters of the phase-averaged GeV spectra. 
The first and second errors represent 1$\sigma$ statistical errors and systematic errors, respectively.}
\label{tab_fermi_fit_allphase}
  \begin{center}
    \begin{tabular}{c c c} \hline
 & \multicolumn{2}{c}{All phase} \\
 & Cutoff-PL & Cutoff-PL + PL \\ \hline
$\Gamma$ & $2.32 \pm 0.02 \pm 0.26$ & $2.31 \pm 0.02 \pm 0.21$ \\
$E_\mathrm{cut}$ (GeV) & $5.3 \pm 0.5 \pm 1.5$ & $4.8 \pm 0.5 \pm 0.4$ \\
$N_0$$^\dagger$ & $4.1 \pm 0.1 \pm 0.4$ & $4.2 \pm 0.1 \pm 0.8$ \\
$\Gamma_{\mathrm{TeV}}$ & -- & $0.8 \pm 0.4 \pm 0.5$ \\
$N_{\mathrm{TeV}}$$^\ddagger$ & -- & $ 1.5 \pm 0.4 \pm 0.6 $ \\ \hline
$TS$ value & 13894.5 & 13929.8 \\ \hline
    \end{tabular}
\end{center}
{\small
$\ast$: Cutoff-PL and PL represent the exponentail cutoff model of Eq~\eqref{eq_model_ls} and the power-law model of Eq~\eqref{eq_model_ls_TeV}, respectively.\\
$\dagger$: Its unit is $10^{-11}~\mathrm{cm^{-2}~s^{-1}~MeV^{-1}}$. \\
$\ddagger$: Its unit is $10^{-15}~\mathrm{cm^{-2}~s^{-1}~MeV^{-1}}$. \\
}
\end{table}

We also analyzed the spectra for the INFC and SUPC orbital phase intervals.
The definition of INFC and SUPC are the same as those used for the \nus analysis (see \S\ref{sec_nustar_ana_allene}), but we adopted the most up-to-date orbital parameters; the orbital period of $3.90608 \pm 0.00010$ days \citep{aragonaorbits2009} and $T_0 = 2455017.08 \pm 0.06$ (HJD) \citep{Sarty2011}
in order to reduce the uncertainty of the orbital phase calculation.
The reference time of \cite{casarespossible2005} used in the X-ray analysis is
2001 February 2, which is at maximum $\sim 19$ years apart from the LAT data.
Considering the error on the orbital period reported in it,
the orbital phase calculation for the LAT data has an uncertainty of $\sim 0.08$ at maximum.
This is comparable to the bin width of the orbital light curves in \S\ref{sec_orbital_light_curve_GeV} and may smear out any fine structure in them.
By using the up-to-date parameters above,
the uncertainty can be reduced to $\lesssim 0.02$ which is accurate enough for our analysis.

As presented in Figure~\ref{fig_fermi_fit}b, 
the SUPC spectrum was well described by a sum of Eq.~\eqref{eq_model_ls} and Eq.~\eqref{eq_model_ls_TeV}.
On the other hand,
the INFC spectrum shows a hump structure around a few GeV.
Although this spectral feature was mentioned in \citet{Hadasch2012},
they modeled the spectrum with a single exponential-cutoff power law.
Here,
we modeled it by adding a power-law component.
Table~\ref{tab_fermi_fit}  describes the best-fitting parameters.
The three-component model improved the $TS$ value by $\Delta TS = 84$ and this yields a chance probability of $1\times10^{-18}$.
Thus, the three-component model is significantly favored.
The cutoff power-law component with $\Gamma \simeq 0.9$ and $E_\mathrm{cut} \simeq 1.4$ GeV
dominates the SED from $\sim 400$ MeV to $\sim 10$ GeV.
The spectrum below $\sim$ 400 MeV is dominated by an additional component,
which is explained by the single power law with a much larger index of $\Gamma \sim 3.0$.

In the SUPC spectrum, the hump structure is hardly seen, and adding a power-law component did not improve the fitting.
This can be interpreted as that
the low-energy component described by the single power law in the INFC changes its spectral shape in the SUPC {\it e.g.}
the photon index becomes smaller or the spectral cutoff becomes non-negligible,
Thus, we fitted the SUPC spectrum by a sum of two exponential-cutoff power-law components and a single power law.
The two exponential-cutoff power laws correspond to the low-energy component and the hump structure seen in INFC,
and the single power law is the low-energy tail of the TeV component.
As a result, we found that the $TS$ value was improved by 49. 
Note that $\Gamma_2$ and $E_{\mathrm{cut},2}$ in Table~\ref{tab_fermi_fit} were fixed to the values obtained from the INFC spectrum because these values were found to be strongly coupled with $\Gamma_1$ and $E_{\mathrm{cut},1}$.

\begin{table*}[!htbp]
\caption{Best-fitting parameters of the GeV spectra in INFC/SUPC. 
The first and second errors represent 1$\sigma$ statistical errors and systematic errors, respectively.}
\label{tab_fermi_fit}
  \begin{center}
    \begin{tabular}{c c c c c } \hline
 & \multicolumn{2}{c}{INFC} & \multicolumn{2}{c}{SUPC} \\
 & Cutoff-PL + PL & PL + Cutoff-PL + PL & Cutoff-PL + PL & Cutoff-PL + Cutoff-PL + PL \\ \hline
$\Gamma_1$ & $2.39 \pm 0.03 \pm 0.44$                & $3.0 \pm 0.2\pm 0.5$  & $2.2 \pm 0.03 \pm 0.21$ & $1.6 \pm 0.1 \pm 0.4$    \\
$E_\mathrm{cut,1}$ (GeV) & $10.6 \pm 2.4 \pm 7.7$    & --                    & $2.8 \pm 0.4 \pm 0.3$   & $0.35 \pm 0.06 \pm 0.06$ \\
$N_{0,1}$$^\ast$ & $ 2.7 \pm 0.1 \pm 0.3$            & $0.9 \pm 0.3 \pm 0.5$ & $6.0 \pm 0.3 \pm 0.6$   & $25.8 \pm 7.7 \pm 10.6$  \\
$\Gamma_2$  & --                                     & $0.9 \pm 0.4 \pm 0.1$ & --                      & (0.9)                  \\
$E_\mathrm{cut,2}$ (GeV) & --                        & $1.4 \pm 0.4 \pm 0.1$ & --                      & (1.4)                  \\
$N_{0,2}$$^\ast$ & --                                & $3.4 \pm 0.5 \pm 0.8$ & --                      & $4.5 \pm 0.3 \pm 0.2$  \\
$\Gamma_{\mathrm{TeV}}$ & $ -0.1 \pm 0.7 \pm 0.6$    & $1.1 \pm 0.6 \pm 1.0$ & $1.6 \pm 0.6 \pm 0.3$   & $1.6 \pm 0.4 \pm 0.1$ \\
$N_{\mathrm{TeV}}$$^\dagger$ & $0.7 \pm 0.5 \pm 0.6$ & $1.9 \pm 0.8 \pm 0.7$ & $2.2 \pm 0.6 \pm 0.1$   & $2.2 \pm 0.6 \pm 0.1$\\ \hline
$TS$ value & 3863.6 & 3948.0 & 10605.9 & 10655.2\\ \hline
    \end{tabular}
\end{center}
{\small
$\ast$: Its unit is $10^{-11}~\mathrm{cm^{-2}~s^{-1}~MeV^{-1}}$. 
$\dagger$: Its unit is $10^{-15}~\mathrm{cm^{-2}~s^{-1}~MeV^{-1}}$. \\
}
\end{table*}

\subsection{Dependence of the Flux on the Orbital Phase}
\label{sec_orbital_light_curve_GeV}

To study the orbital variations of the gamma-ray flux at different energies,
we divided the 11 years of LAT data into 10 subsets,
each covering an orbital phase of $\Delta\phi = 0.1$.
Then, we measured the flux in several energy bands from each subset,
by modeling \ls with a power-law function in each energy band.
Here the energy intervals were defined as three bins per decade in logarithmic scale.
The spectral parameters except for \ls were fixed to the values obtained in the orbit-averaged spectral analysis.

\begin{figure}[!htbp]
    \begin{center}
    \includegraphics[width = 7.5 cm]{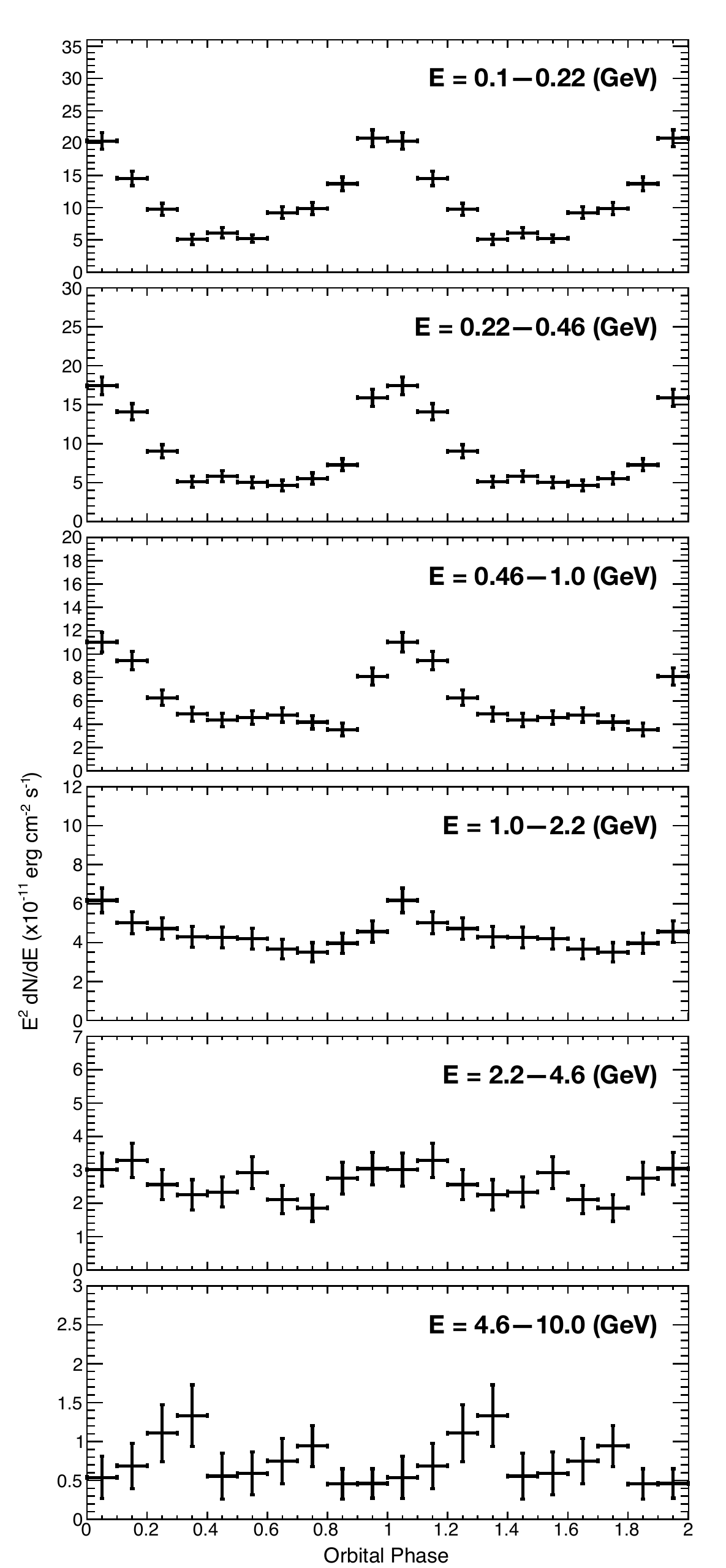}
    \end{center}
\caption{The orbital light curves in the GeV gamma-ray band in different energy intervals.}
\label{fig_fit_fermi_lc_phasedep}
\end{figure}

Figure~\ref{fig_fit_fermi_lc_phasedep} shows the orbital light curves derived in different energy intervals.
From 100 MeV to 1 GeV,
the light curves have a strong peak around $\phi \sim 0.0$, where the compact object is closest to the companion star.
The rise time of the peak is longer at 100--220 MeV than at 460--1000 MeV.
Interestingly, above 1 GeV,
the orbital modulation becomes considerably weaker, as also mentioned in \cite{Chang2016}.

\begin{figure}[!htbp]
\begin{center}
\includegraphics[width = 8.5 cm]{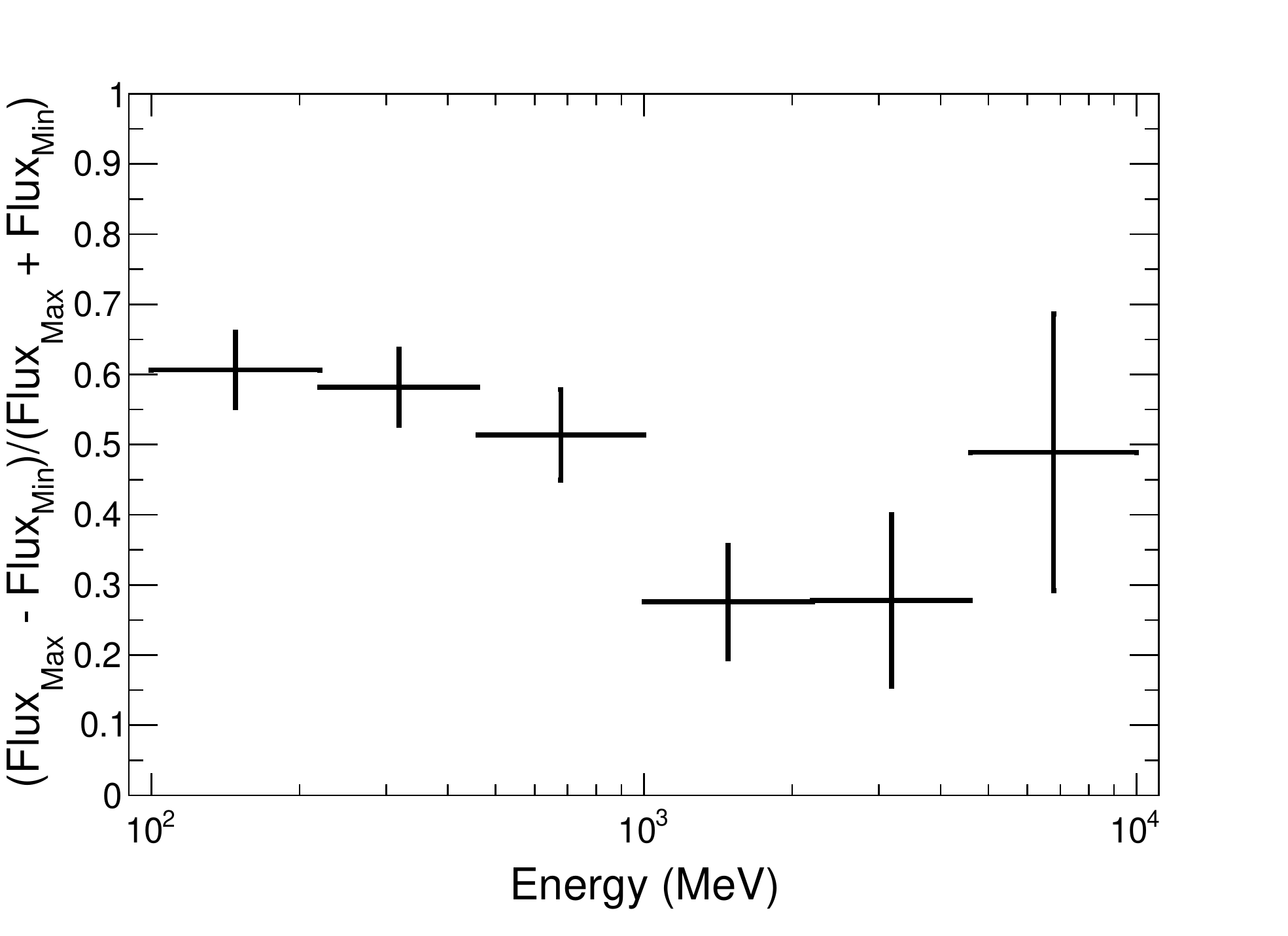}
\end{center}
\caption{Energy dependence of $\xi$, defined by Eq.~\ref{eq_ratio}.}
\label{fig_fit_fermi_phase_fluxratio}
\end{figure}

In order to evaluate this property more quantitatively,
we calculate a quantity defined as
\begin{equation}
\label{eq_ratio}
\xi = (F_\mathrm{max} - F_\mathrm{min}) / (F_\mathrm{max} + F_\mathrm{min})~,
\end{equation}
where $F_\mathrm{max}$ and $F_\mathrm{min}$ are the maximum and minimum flux in the light curves, respectively.
This $\xi$ is interpreted approximately as the fraction of the variable component over the total flux.
As shown in Figure~\ref{fig_fit_fermi_phase_fluxratio}, the significant decrease of $\xi$ is observed at $\sim$ 1 GeV;
while the fraction of the variable component is $\sim$ 60\% at $\sim$ 300 MeV,
it decreases to $\sim$ 30\% in 1--5 GeV.
Considering together the spectral change at $\sim$ 1 GeV (see Figure~\ref{fig_fermi_fit}),
this result suggests that there are two spectral components below 10 GeV;
one is dominant below $\sim$ 1 GeV with a steep photon index in INFC while it becomes harder with a cut-off at $\sim$ 400 MeV in SUPC,
and the other is dominant above $\sim$ 1 GeV and nearly independent of the orbital phase.

\subsection{Search for Time Variability}

\begin{figure}[!htbp]
\begin{center}
\includegraphics[width = 8.5 cm]{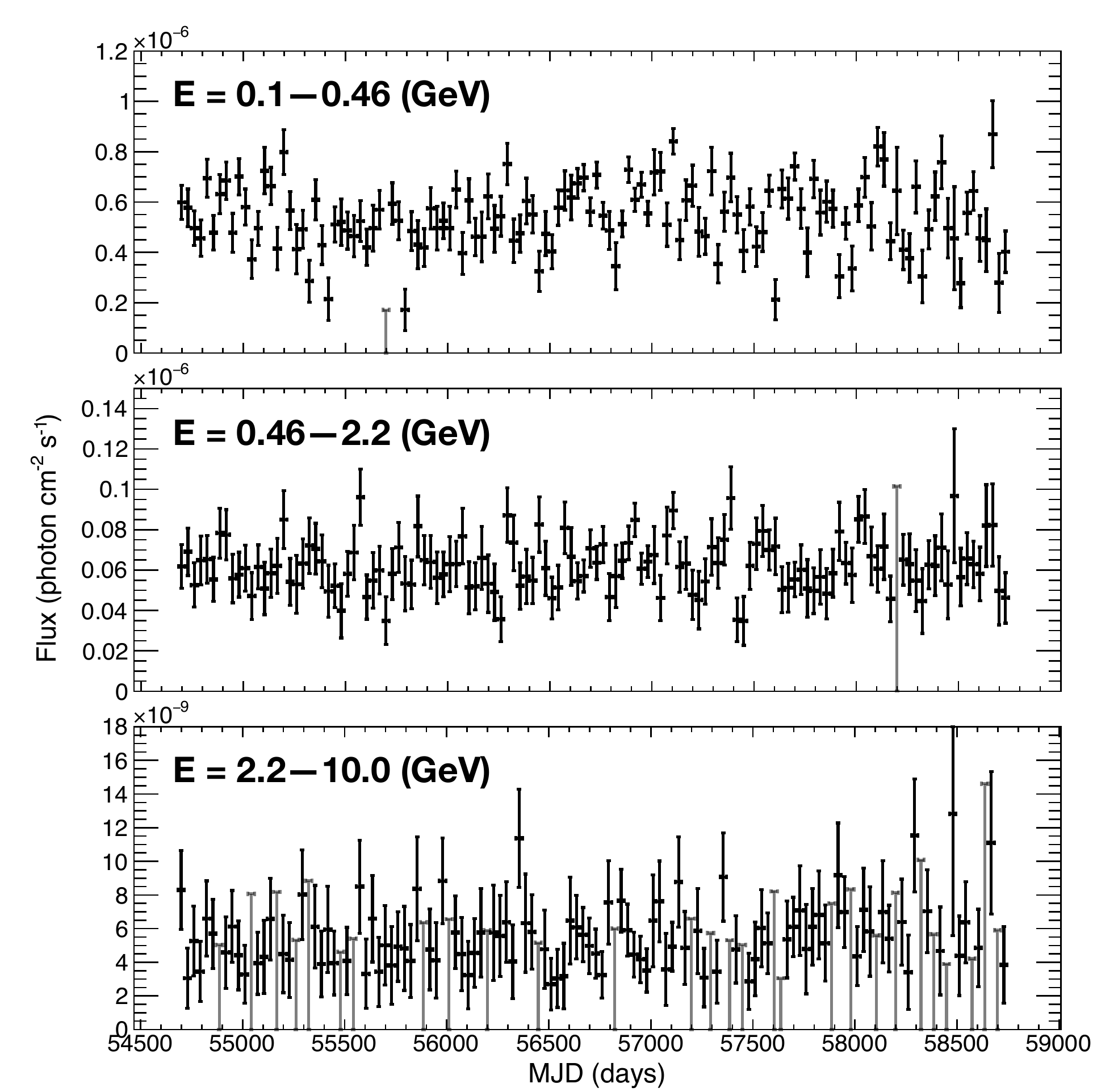}
\end{center}
\caption{Light curves of \ls with a bin width of 31.25 days in the different three energy bands. The gray points represent the upper limit of 95\% confidence level.}
\label{fig_lc_gev_1month}
\end{figure}

\begin{figure}[!htbp]
\begin{center}
\includegraphics[width = 8.5 cm]{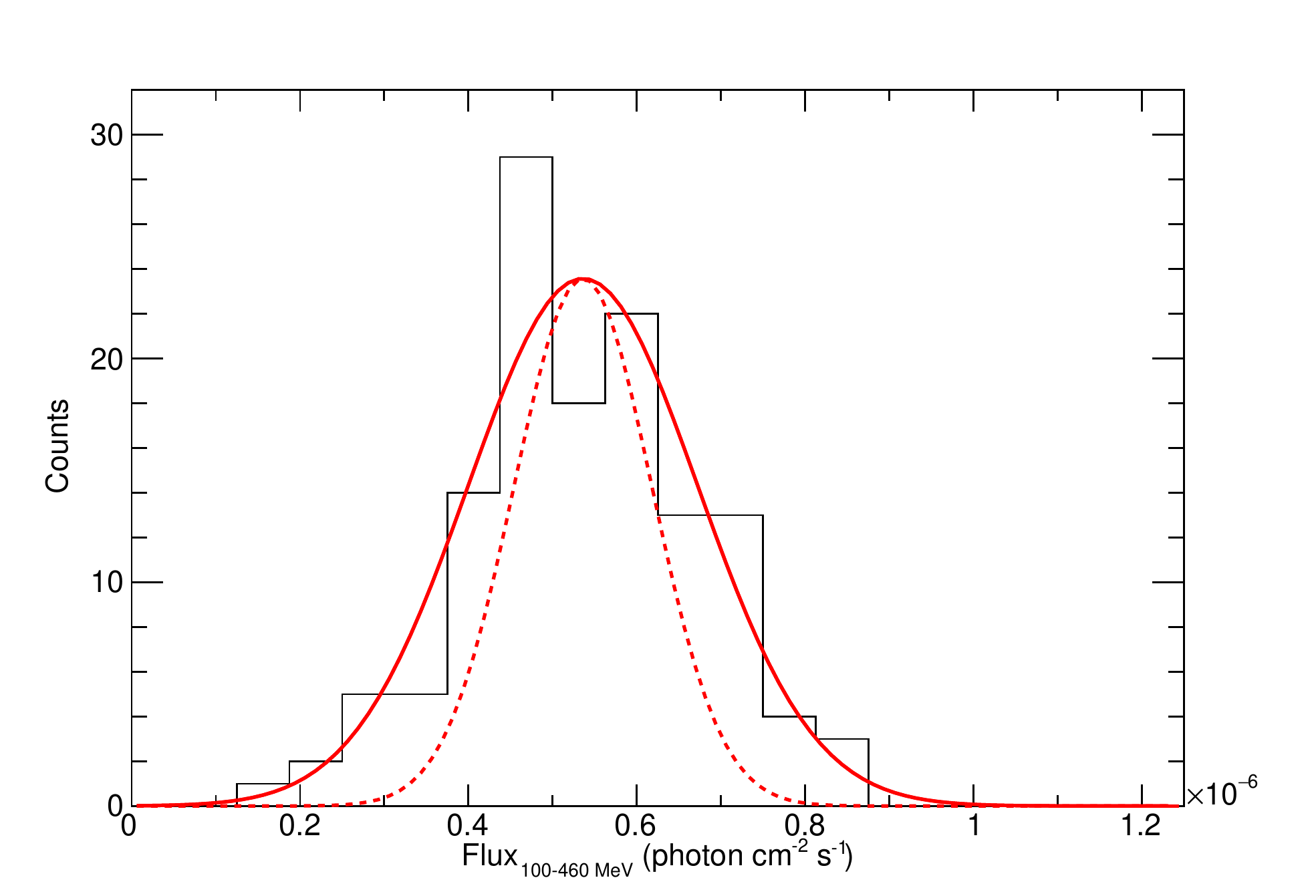}
\end{center}
\caption{An histogram of 100--460 MeV photon fluxes in 31.25 days interval.
The solid red line represents the best-fit curve of a Gaussian function.
The dotted one is a Gaussian function with the same normalization of the solid line but its standard deviation is an average of statistical errors of the obtained photon flux.}
\label{fig_hist_all_E_100_460_v2}
\end{figure}

In order to investigate long-term variability,
we produced the source light-curve with a bin width of about one month.
We fitted each 31.25 days interval which is 8 times of the orbital period,
by leaving free all of the parameters of \ls and PKS1830-211.
The other spectral parameters were again fixed to the best-fitting values in the orbit-averaged spectral analysis.
Then we measured the flux of \ls using a power-law function in three energy intervals; 100--460 MeV, 460 MeV--2.2 GeV, and 2.2--10 GeV. 
The obtained light curves are shown in Figure~\ref{fig_lc_gev_1month}.
In the 11 years of \fermi observation, no flaring activity was found.
In 100--460 MeV,
the average and standard deviation of the flux were obtained as
$(5.4 \pm 0.1) \times 10^{-7}$ photon cm$^{-2}$ s$^{-1}$ and
$(1.4 \pm 0.1) \times 10^{-7}$ photon cm$^{-2}$ s$^{-1}$, respectively.
As shown in Figure~\ref{fig_hist_all_E_100_460_v2}, this standard deviation was larger than an average of statistical errors ($0.8 \times 10^{-7}$ photon cm$^{-2}$ s$^{-1}$). 
Furthermore, when the light curve was fitted by a constant value,
the chi-square of the fitting was 375.0 with 128 degrees of freedom
and its corresponding chance probability is $4\times10^{-26}$.
Thus, the obtained flux deviation would indicate intrinsic source variability which is about 20\% of the averaged flux.
Above 460 MeV,
the flux fluctuation is consistent with the statistical errors,
and the variability is less than 20\% and 40\% in 460 MeV--2.2 GeV and 2.2--10 GeV, respectively.

\section{Discussion}
\label{sec_discussion}
In this work, we studied the spectrum of \ls using the \nus and \fermi LAT data.
Even with the high sensitivity of \nus,
we found that the hard X-ray spectrum of \ls is well explained by a single power-law component up to 78 keV.
We also analyzed the 11 years of \fermi LAT data, and obtained the spectrum in the GeV band with highest available statistics.
We confirmed the spectral hump at $\sim$ 2 GeV which was suggested by \cite{Hadasch2012},
and revealed that \ls has two spectral components in the GeV band.

\subsection{A possibility of a new spectral component above 100 keV}
\label{sec_discussion_MeV}

\begin{figure*}[!htbp]
    \begin{center}
    \includegraphics[width = 15 cm]{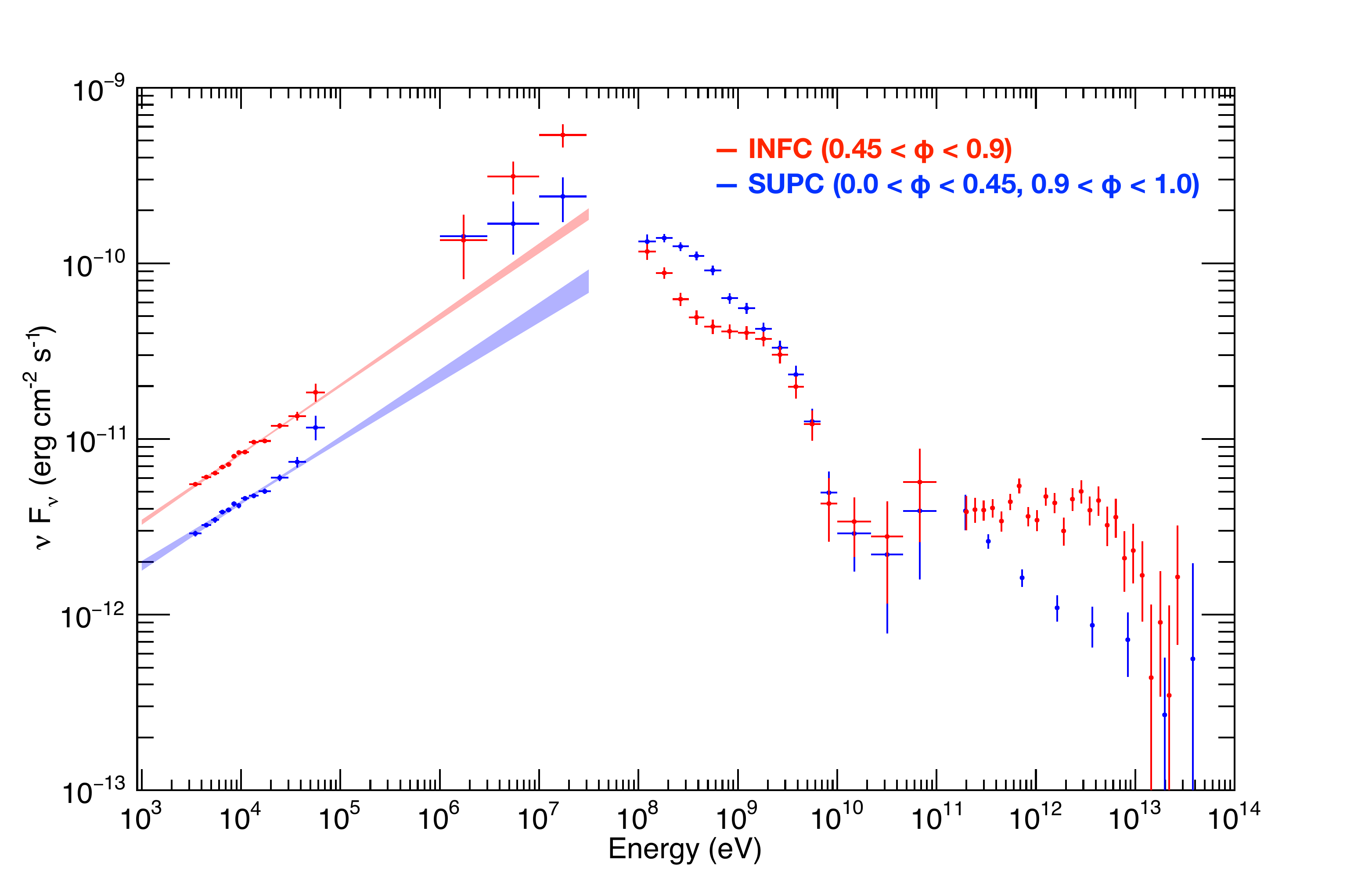}
    \end{center}
\caption{An updated SED of \ls. 
The red and blue points indicate the data at INFC and SUPC phases, respectively.
The flux points in $10^3$--$10^5$ eV and $10^8$--$10^{11}$ eV are obtained in this work.
The MeV and TeV data are taken from \citet{collmarls2014} and \citet{hess2006}, respectively.
The colored region is extrapolation of the X-ray emission to the MeV band using the power-law model with the parameters in Table~\ref{tab_spectrum_3_70_keV_nustar}.
}
\label{fig_sed_ls5039}
\end{figure*}

Figure~\ref{fig_sed_ls5039} shows the SED of \ls,
which has been updated from that of \citet{collmarls2014} incorporating our \nus and \fermi results.
Using the power-law model with the parameters shown in Table~\ref{tab_spectrum_3_70_keV_nustar},
we extrapolated the X-ray emission to the MeV band,
which is shown as the colored region in Figure~\ref{fig_sed_ls5039}.
It is significantly smaller than the result of \comp at both SUPC and INFC,
and thus the extrapolation of the X-ray continuum does not smoothly connect with the \comp fluxes.
Since the hard X-ray spectra obtained with \nus were successfully described
up to 78 keV by the single power-law model,
this indicates that another spectral component should emerge above $\sim$ 100 keV.
In addition, the hard X-ray spectrum at the SUPC shows a sign of the spectral hardening above $\sim$ 50 keV, even though the statistical error is large.

While the obtained spectrum suggests another spectral component above $\sim$ 100 keV,
we should note two cautions on the interpretation of the \comp data.
One is that the \comp and \nus data were apart by $\sim$ 20 years and 
we cannot exclude a possibility that the difference between the two data sets is caused by a long-term variability in the MeV band.
The other is that the angular resolution of \comp is 1.7--4.4 degrees and the obtained flux might be overestimated due to the contamination from unresolved 
sources close to \ls.
This contamination can be estimated as one-third of the derived value at most by
assuming that the offset in the orbital light curve is completely due to the unresolved sources \citep{collmarls2014}.
Even with this extreme case, the MeV gamma-ray flux at INFC is still larger than the extrapolation of the \nus data.
\cite{Falanga2021} and \cite{Malizia_intergral} also reported the spectral hardening with the analysis of \integralsat data.

\begin{figure*}[!htbp]
    \begin{center}
    \includegraphics[width = 8.5 cm]{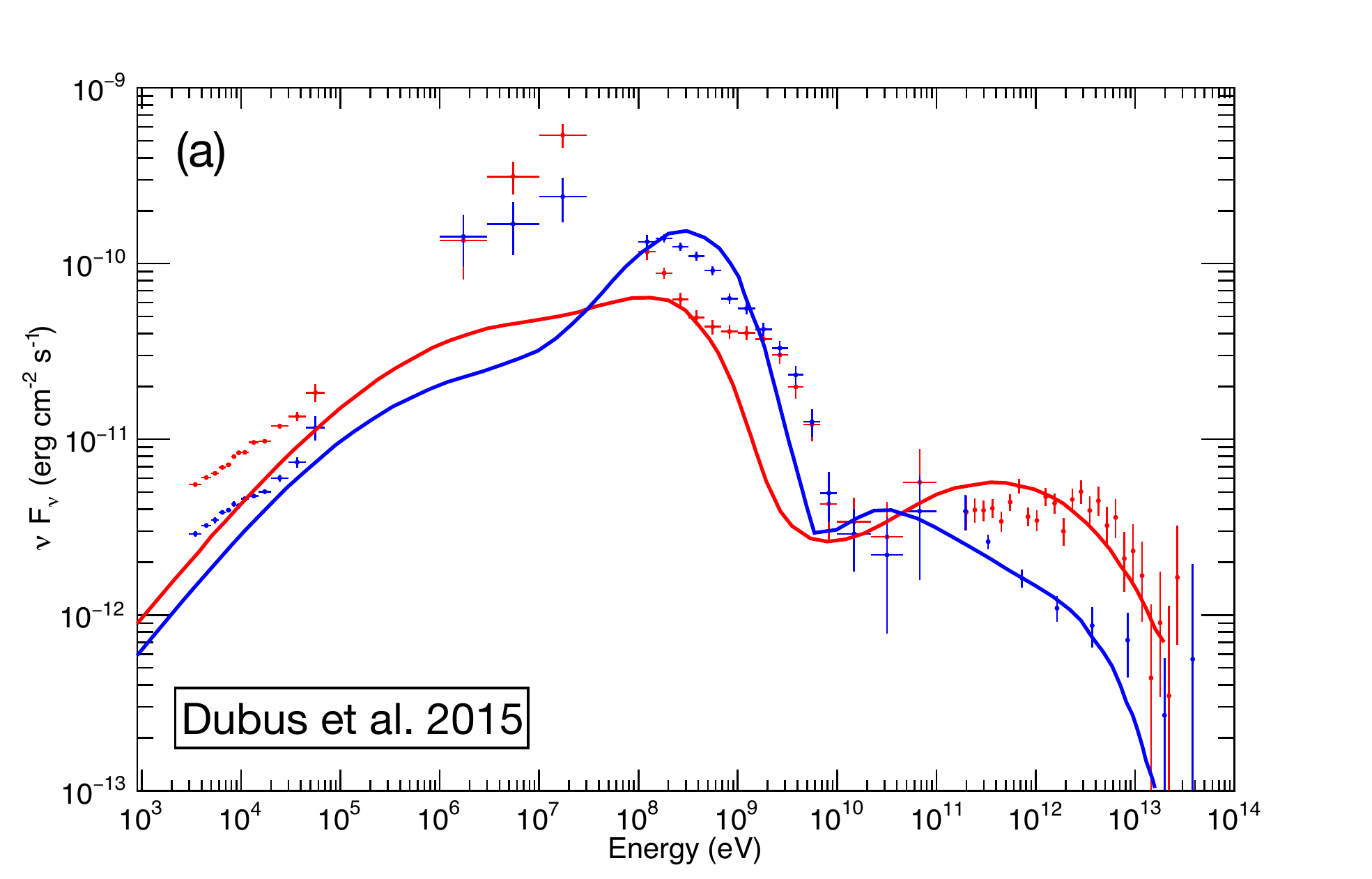}
    \includegraphics[width = 8.5 cm]{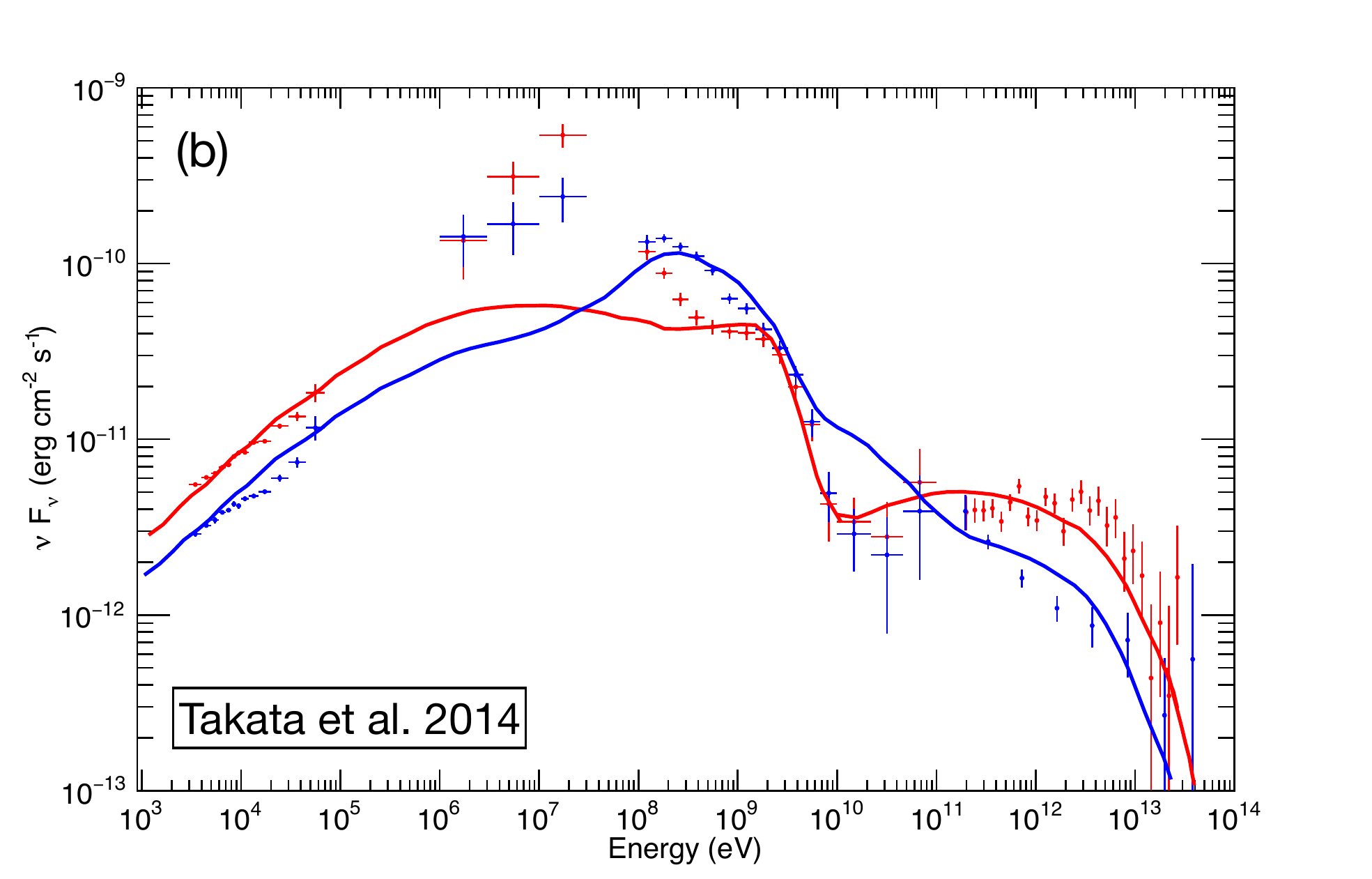}
    \includegraphics[width = 8.5 cm]{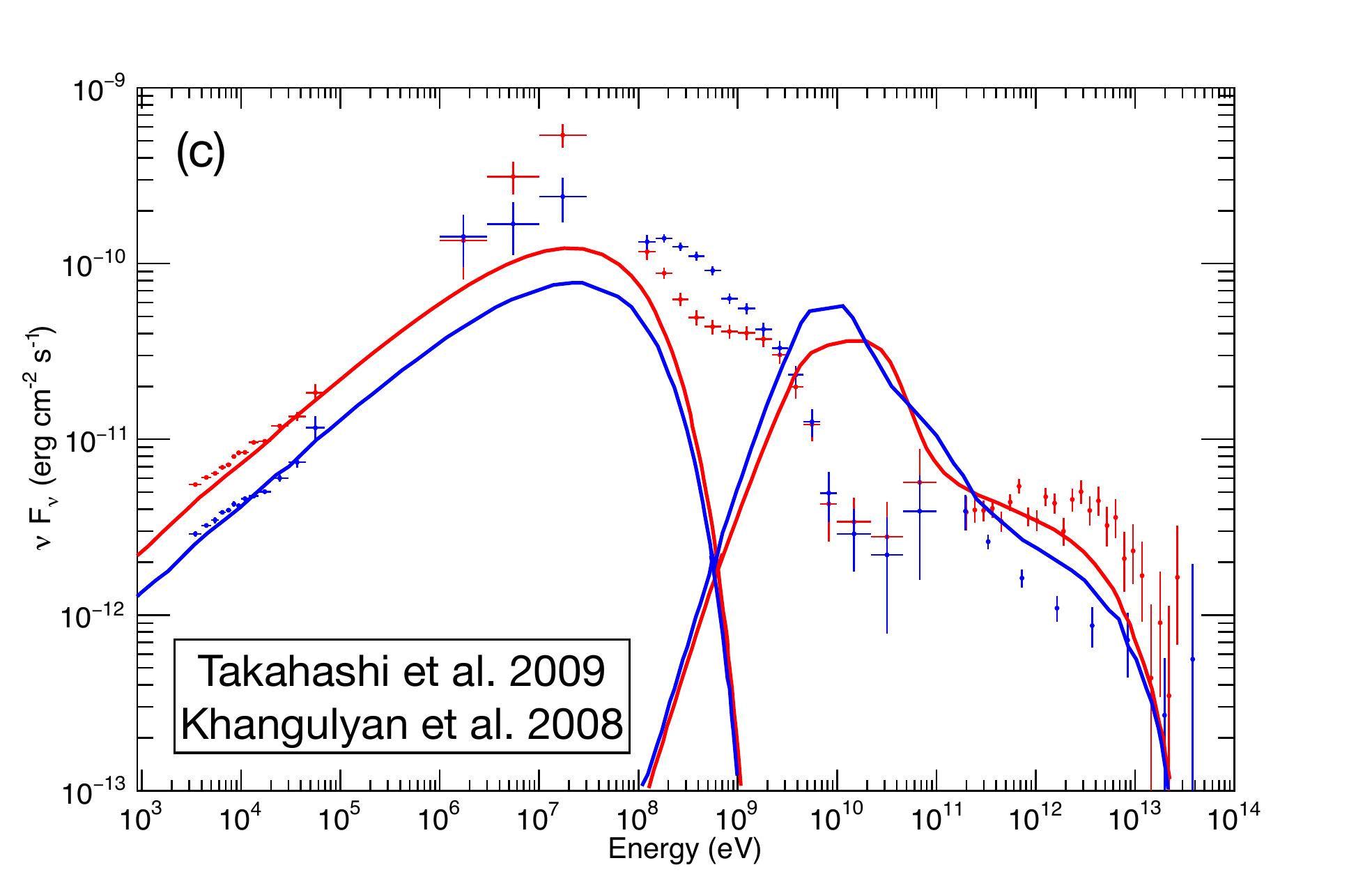}
    \includegraphics[width = 8.5 cm]{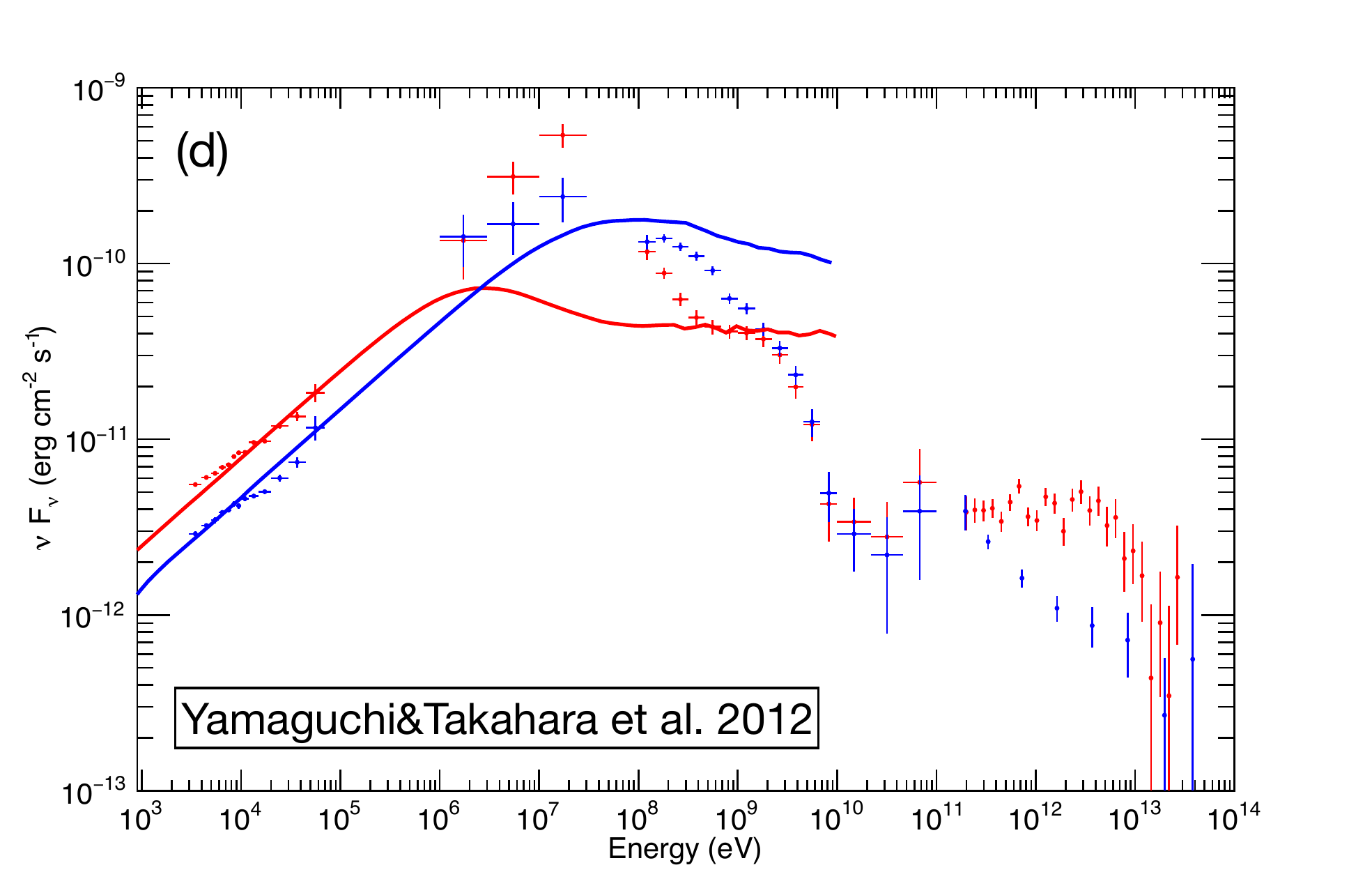}
    \end{center}
\caption{The comparison between the updated SED and proposed models.
(a) A pulsar wind scenario in which a narrow Maxwellian-distributed electrons are assumed for the GeV emission \citep{dubus2015}.
(b) A pulsar wind scenario in which the GeV emission is interpreted as the outer gap emission near the neutron star \citep{takata2014}.
(c) A microquasar model in which the gamma-ray emission is produced in jet-like relativistic outflows \citep{khangulyan2008}.
(d) A two-component inverse Compton scattering model in which the X-ray emission is explained by the inverse Compton emission of low energy electrons with a gamma factor of $< 10^3$ \citep{Yamaguchi2012}.}
\label{fig_model_comparsion_sed_ls5039}
\end{figure*}

The presence of an additional non-thermal component above 100 keV is also supported by some theoretical arguments.
\cite{dubus2015} studied a spectral model of \ls based on the pulsar winds scenario using numerical relativistic hydrodynamics. 
As quoted in Figure~\ref{fig_model_comparsion_sed_ls5039}a,
they found that the simulated  X-ray and MeV gamma-ray fluxes are significantly below the measured flux.
They further speculated that a better agreement may be achieved if the source harbors an additional production site with a stronger magnetic field.
Other spectral models including 
the pulsar wind scenario \citep[Figure~\ref{fig_model_comparsion_sed_ls5039}b;][]{takata2014}, 
the microquasar scenario \citep[Figure~\ref{fig_model_comparsion_sed_ls5039}c;][]{khangulyan2008}, 
and the two-component inverse Compton emission model \citep[Figure~\ref{fig_model_comparsion_sed_ls5039}d;][]{Yamaguchi2012},
also fail to explain the gamma-ray emission from $\sim$ 1 MeV to $\sim$ 400 MeV obtained with \comp and \fermi.

As discussed in \cite{dubus2015}, the most plausible interpretation of the MeV component is that it is synchroton emission from region(s) of strong magnetic fields as discussed below.
If the maximum energy of electrons $E_e,_\mathrm{max}$ is determined by the balance between the acceleration and synchrotron losses, then
$E_e,_\mathrm{max}$ is given as \citep{khangulyan2008}
\begin{eqnarray}
E_e,_\mathrm{max} = 60~\mathrm{TeV}~\left(\frac{B}{1~\mathrm{G}}\right)^{-0.5} \eta^{-0.5}~,
\end{eqnarray}
where $B$ is the magnetic field strength in the acceleration region and $\eta$ is a parameter which characterizes acceleration efficiency \citep[see also][]{Aharonian2002}.
The SED of synchrotron emission from electrons with $E_{e,}{}_\mathrm{max}$ peaks at \(h\nu \simeq 236\, \eta^{-1} \,\rm MeV\) \citep{2012SSRv..173..341A}.
Since the emission from \ls has the spectral peak at 20-30 MeV as shown in Figure~\ref{fig_sed_ls5039},
$\eta$ is estimated as $<$10.
Such a small $\eta$, {\it i.e.} an extreme efficient acceleration, was also suggested in \citet{takahashistudy2009}.
Because diffusive shock acceleration (DSA) yields $\eta > 10$ even in the Bohm limit,
the obtained small $\eta$ suggests that 
the extreme efficient particle acceleration operating in \ls is 
different from DSA.

\begin{figure}[!htbp]
    \begin{center}
    \includegraphics[width = 8.5 cm]{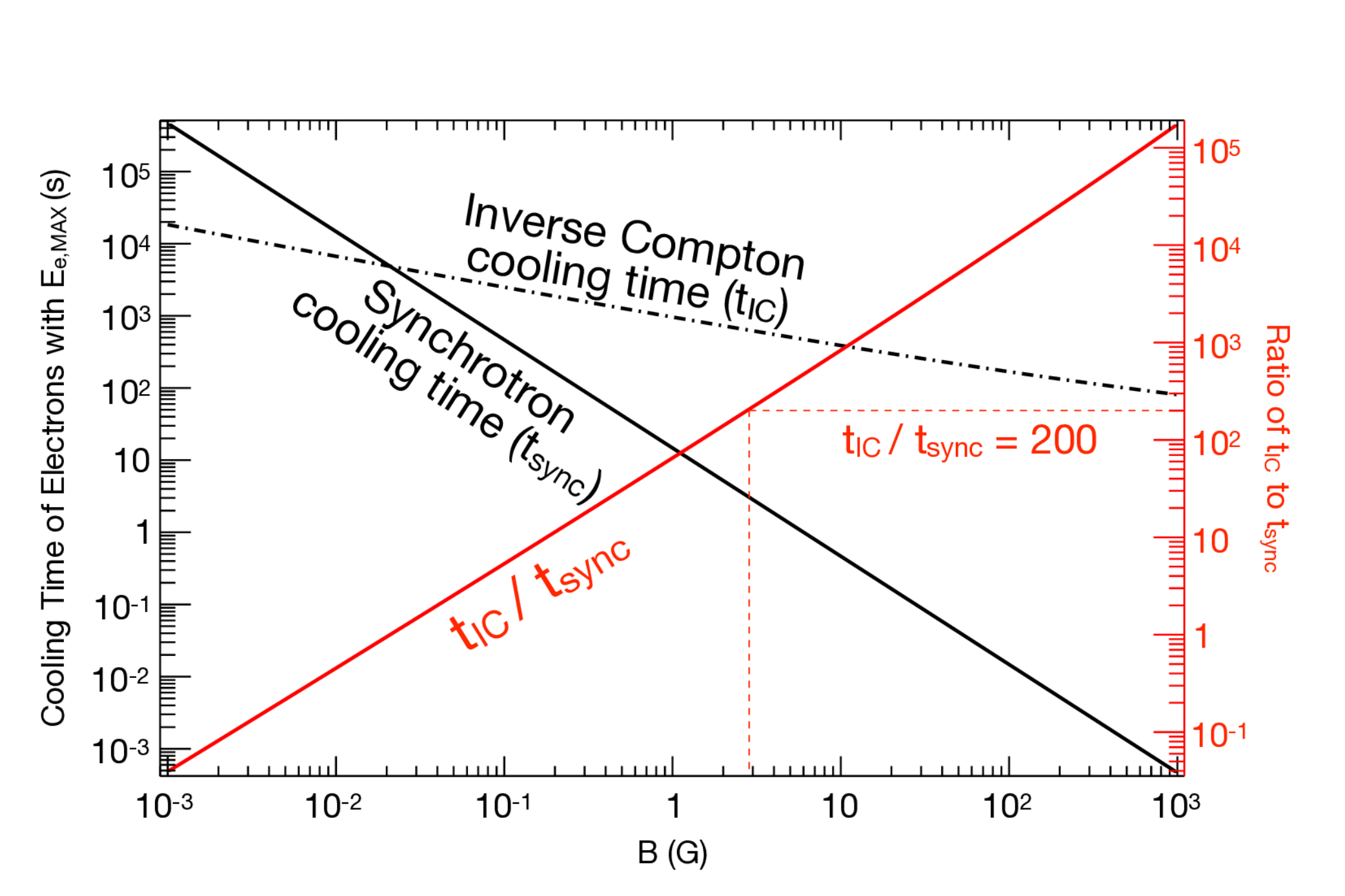}
    \end{center}
\caption{The cooling times of synchrotron and inverse Compton losses for electrons with $E_{e,\mathrm{MAX}}$. Here $\eta = 5$ is assumed. For the calculation of $t_\mathrm{IC}$, the approximation in \cite{khangulyan2014} was used by employing 
the values of the parameters ($kT_\star, R_\star, D_{\mathrm{sep}}$) as described in Table~\ref{tab_parameters}.}
\label{fig_cooling_time}
\end{figure}

The accelerated electrons also produce TeV gamma rays 
by boosting up ultraviolet photons from the companion star via inverse Compton scattering.
In Figure~\ref{fig_cooling_time} we show the cooling time of synchrotron ($t_\mathrm{sync}$) and inverse Compton ($t_\mathrm{IC}$) losses for electrons of energy $E_{e,\mathrm{max}}$ assuming \(\eta=5\).
Since the luminosities from these radiation processes $L_\mathrm{sync}$ and $L_\mathrm{IC}$ are proportional to the inverse of their cooling times,
we find $L_\mathrm{sync}/L_\mathrm{IC} \simeq t_\mathrm{IC}/t_\mathrm{sync}$.
If we interpret the observed MeV gamma-ray emission as the synchrotron process,
then corresponding inverse Compton emission should be comparable to or weaker than the observed TeV gamma-ray emission.
Thus, from the SED in Figure~\ref{fig_sed_ls5039},
we can obtain $L_\mathrm{sync}/L_\mathrm{IC} \gtrsim 200$.
Then, the cooling times shown in Figure~\ref{fig_cooling_time} implies that the magnetic field at the production site is higher than $\sim 3$ G, unless it is located significantly further away from the optical companion than the system size.
\cite{yoneda2020} proposed that \ls harbors a magnetar, based on the hints of 9 sec pulsation in hard X-rays.
This magnetar hypothesis may be one of scenarios that can explain the MeV emission based on this synchrotron interpretation. 

An alternative interpretation is that the MeV component is produced via inverse Compton scattering.
In this case,
the characteristic energy of the electrons responsible for the MeV emission is $E_e \simeq 1$ GeV.
If we assume that the pressure of these GeV electrons is equal to the ram pressure of the stellar wind close to the compact star,
we can estimate the size scale $R$ of the emission region as follows.
When we assumed that the production site is close to the compact object,
the cooling time, $\tau_e$, of these electrons through their invese Compton scattering off the optical photons from the optical companion is estimated as about \(10^2\)~s \citep{khangulyan2008}.
The system parameters assumed here are described in Table~\ref{tab_parameters}.
Then, the total amount of electron energy in the MeV emission region is estimated as $E_{e,\rm total} = L_\mathrm{MeV} \times \tau_e \simeq 10^{38}~\mathrm{erg}$.
Since the pressure created by the GeV electrons is $P \sim E_{e,\rm total}/(4\pi R^3)$,
and the ram pressure of the stellar wind is $P = (\dot{M} v_{wind})/(4\pi D_{\mathrm{sep}}^2) \sim 10^2~\mathrm{erg~cm^{-3}}$,
we obtain an estimate as $R \sim 10 ~\mathrm{light~sec}$.
This value is comparable to the size scale of the colliding shock in the pulsar wind scenarios \citep{dubus2015}.
If there is a sub-population of GeV electrons in the shock region in addition to TeV electrons,
then the MeV component may be explained by the inverse Compton emission. 
However, in this case, we expect a different orbital phase dependence: the IC flux from the SUPC orbital phase interval should be higher than that from INFC interval.
Therefore, this alternative is less likely than the synchrotron scenario. 

\begin{table}[!htbp]
  \begin{center}
  \caption{The parameter values assumed in the inverse Compton scenario, referring to \cite{casarespossible2005}.}
\label{tab_parameters}
    \begin{tabular}{c c} 
parameter & value \\ \hline
$kT_\star$ & 3.3 eV\\
$R_\star$ & 9.3 $R_\odot$\\
$\dot{M}$ & $10^{-6}~M_\odot~\mathrm{year}^{-1}$ \\
$v_{wind}$ & $2000~\mathrm{km~s^{-1}}$\\
$D_{\mathrm{sep}}~^*$ & 100 light-sec \\
\hline
    \end{tabular}
\end{center}
{\small $*$: It represents the binary separation.}
\end{table}

\subsection{The origin of the spectral component from 1 GeV to 10 GeV}
\label{sec_discussion_GeV}

The \fermi 11 year observations show that
there are likely two spectral components below 10 GeV,
and the higher-energy component has weak dependence on the orbital phase.
The flux modulation $\xi$, defined in Eq.~\ref{eq_ratio},
is at most 20--30\% in 1--5 GeV.
Therefore,
the spectral component in $1$--$10$ GeV
must be produced by a process that depends weakly on the physical parameters related to the binary motion, {\it e.g.}, the binary separation and the IC scattering angle.

Among several scenarios proposed for the GeV emission,
\citet{dubus2015} assumed that
it is produced by electrons distributed with a narrow Maxwellian in the shock region (Figure~\ref{fig_model_comparsion_sed_ls5039}a).
Existence of such particle populations is supported by some particle-in-cell simulations, 
which show that the relativistic shock produces Maxwellian-distributed electrons in addition to power-law-distributed ones \citep{Sironi2009, Sironi2011}. 
In this scenario, the GeV gamma rays are produced via inverse Compton scattering by the Maxwellian-distributed electrons.
However,
this scenario predicts that the GeV flux varies significantly ($\sim$ 70\%) depending on the orbital phase,
because the scattering angle and the power injected as the Maxwellian-distributed electrons strongly depend on the orbital phase.
Consequently, the predicted light curve is inconsistent with that obtained with \fermi.

As another possibility, the GeV electrons might be produced
when TeV gamma rays are converted into electron-positron pairs via gamma-gamma absorption.
Then, they would emit the GeV gamma rays via inverse Compton scattering.
In order to explain the GeV emission of \ls in this scenario,
at least 90\% of the TeV gamma rays should be absorbed
because the flux of the GeV component is an order of magnitude higher than that of the TeV component.
However, such a strong absorption is difficult to achieve in \ls \citep{Dubus2006}.
Alternatively, GeV gamma rays might be produced via pion decay process
if protons are accelerated in the shock region
and they could interact with unaccelerated protons in the stellar winds or the interstellar medium.
However, 
if the target proton density is determined by the stellar wind around the compact object,
the cooling time is estimated as $\sim 10^5$ s \citep{felixblackbook},
which would require energy injections much larger than those of the leptonic models and would make it even more challenging to explain the observed flux level \citep[see e.g.][for review]{boschramon2009}.

\citet{takata2014} interpreted the GeV component as the emission from the outer gap in a pulsar's magnetosphere (Figure~\ref{fig_model_comparsion_sed_ls5039}b).
In this scenario,
particles are accelerated in a region where the charge density changes sign near the light cylinder.
As the emission is produced very close to the pulsar,
its gamma-ray flux would not depend on the orbital motion.
Because the flux of the outer gap emission is considered to be constant over the entire orbit,
this scenario can qualitatively explain the small orbital dependence of the GeV emission.

\subsection{Summary of the spectral components of \ls}
In the above subsections,
we discussed the properties and origins of the spectral components in the MeV and GeV bands.
Summarizing them,
we conclude that \ls has at least four spectral components from X-ray to TeV band:
(1) an X-ray component up to $\sim$ 100 keV, 
(2) a spectral component which is dominant above $\sim$ 100 keV and peaks at a few tens of 10 MeV (\S\ref{sec_discussion_MeV}),
(3) a GeV gamma-ray component which is almost independent of the orbital motion (\S\ref{sec_discussion_GeV}),
and (4) a TeV gamma-ray component. 
Components (1) and (4) can be interpreted as the synchrotron and inverse Compton emission respectively,
produced by TeV electrons accelerated at the pulsar wind termination shocks \citep{dubus2015,takata2014}. 
On the other hand, the origins of components (2) and (3) are unclear at this moment.
We discussed that component (2) can be interpreted in two ways; synchrotron emission in strong magnetic fields or inverse Compton emission by GeV electrons.
Several scenarios have been proposed for the GeV emission,
but we suggest that the weak orbital phase dependence of component (3) can be taken as a hint on its magnetosphere origin.

In addition to the above spectral components, the emission from 100 MeV to $\sim$ 400 MeV, described by the steep power law in INFC and the cut-off power law in SUPC (see \S\ref{sec_fermi_ana}),
still remains to be explained.
Due to lack of 30--100 MeV data in the SED,
it is difficult to conclude whether this component is a high-energy tail of component (2) or not.
The orbital modulation of this component would be a clue to reveal its origin.
\citet{Chang2016} reported that the orbital phase of the flux peak shifts from $\phi \sim 0.7$ at $\sim$ 10 MeV to $\phi \sim 0.0$ at $\sim 200$ MeV.
This might suggest that the component from 100 MeV to $\sim$ 400 MeV is different from component (2).
In this case, \ls has five spectral components.

\subsection{The sharp feature observed in previous X-ray observations}
\cite{kishishitalongterm2009} detected a sharp X-ray flux increase around the inferior conjunction in the \chandra and \suzaku observation \citep[see Figure~2 of][]{kishishitalongterm2009}.
Surprisingly, the feature with a duration of $\sim$ only 15 ks was observed at the same orbital phase in the two different observations which are 3 years apart.
This phenomenon is difficult to explain in either the pulsar wind model or the microquasar model,
because in these models fast flux variability is considered to result from multiple stochastic fluctuations in hydrodynamics in general.
Thus, the reproducible fast X-ray brightening may have a hint on understanding the acceleration process in \ls.

In the present \nus observation,
the sharp feature was not confirmed (Figure~\ref{fig_lightcurve_hardX}).
This result can be interpreted as:
the sharp feature is generated at the same orbital phase in every orbits but only in the soft X-ray band ($<$ 3 keV).
Note that
it is pointed out that the previous result of the \chandra\ observation has large systematic uncertainties due to the pileup effect  \citep[$>$ 20\%,][]{readeep2011}.
Also, it covers a small fraction of the orbit, and it is also possible to interpret the result from \chandra as the increase of overall X-ray emission, as we observed $\sim$ 10\% flux variability below 10 keV around the INFC.
Thus, future soft X-ray observations over the entire orbit are essential to confirm the reproducibility of the fast X-ray brightening.

\section{Conclusion}
\label{sec_conclusion}
In this work, we analyzed the \nus and \fermi LAT data of the gamma-ray binary system \ls.
Its spectrum in the hard X-ray band is described well with a single power-law component up to 78 keV.
Comparing with the previous observation of \suzaku,
the flux in the range of 3--10 keV varies by 10--20\% around the inferior conjunction between the two observations.
Furthermore,
we did not find a small spike which was observed at $\phi = 0.70$ by \citet{kishishitalongterm2009}.
These results suggest that the emission around the inferior conjunction varies slightly orbit-by-orbit.

We also analyzed 11 years of observations by \fermi LAT.
The spectrum averaged over the orbital phase is well described with a sum of a cut-off power-law component and power-law one.
The power-law component corresponds to the low-energy tail of the TeV component, and it was
significantly detected from the \fermi LAT data for the first time.
Furthermore,
the spectrum around INFC shows a clear hump at $\sim$ 2 GeV,
and below 10 GeV it is well explained by the sum of two components:
a steep power law with a photon index of $\sim 3.0$, and
an exponential cutoff power law with a photon index of $\sim 0.9$ and a cutoff energy of $\sim 1.4$ GeV.
Although the hump structure is not seen clearly in the SUPC spectrum,
including this component also improved its spectral fitting.
The orbital dependence of the GeV gamma-ray flux was also investigated.
While the fraction of the variable component is $\sim$ 60\% at around 300 MeV,
it decreases to $\sim$ 30\% from 1 GeV to 5 GeV.
On the basis of these results, we suggest that there are two spectral components from 100 MeV to 10 GeV.

Combining these results, we updated the SED of \ls and conclude that
\ls has at least four spectral components in its broad-band spectrum:
(1) a X-ray component up to $\sim$ 100 keV, (2) a spectral component which is dominant above 100 keV up to a few tens of 10 MeV, (3) a GeV gamma-ray component which is almost independent of the orbital motion,
and (4) a TeV gamma-ray component. 
While components (1) and (4) can be produced from the same population of electrons at the pulsar wind termination shock,
the nature and counterparts of components (2) and (3) remains uncertain. 
Furthermore, the emission from 100 MeV to $\sim$ 400 MeV possibly includes another spectral component.
Further theoretical works and observations are needed to reveal the origin of these spectral components.

\acknowledgments
The authors would like to thank Jian Li, David Thompson and Gu$\mathrm{\eth}$laugur J\'{o}hannesson for useful comments. 
H.Y. acknowledges the support of the Advanced Leading Graduate Course for Photon Science (ALPS).
We acknowledge support from JSPS KAKENHI grant numbers 16H02170, 16H02198, 17H02866, 17J04145, 18H01246, 18H05463, 18H03722, 18K03694, 20H00153 and 20K22355.

The \textit{Fermi} LAT Collaboration acknowledges generous ongoing support
from a number of agencies and institutes that have supported both the
development and the operation of the LAT as well as scientific data analysis.
These include the National Aeronautics and Space Administration and the
Department of Energy in the United States, the Commissariat \`a l'Energie Atomique
and the Centre National de la Recherche Scientifique / Institut National de Physique
Nucl\'eaire et de Physique des Particules in France, the Agenzia Spaziale Italiana
and the Istituto Nazionale di Fisica Nucleare in Italy, the Ministry of Education,
Culture, Sports, Science and Technology (MEXT), High Energy Accelerator Research
Organization (KEK) and Japan Aerospace Exploration Agency (JAXA) in Japan, and
the K.~A.~Wallenberg Foundation, the Swedish Research Council and the
Swedish National Space Board in Sweden.
 
Additional support for science analysis during the operations phase is gratefully
acknowledged from the Istituto Nazionale di Astrofisica in Italy and the Centre
National d'\'Etudes Spatiales in France. This work performed in part under DOE
Contract DE-AC02-76SF00515.

\appendix

\section{Spectral Analysis of the \suzaku and \nus Data in 10--30 keV}
\label{ap_hxd}
\begin{figure}[!htbp]
\begin{center}
\includegraphics[width = 8.5 cm]{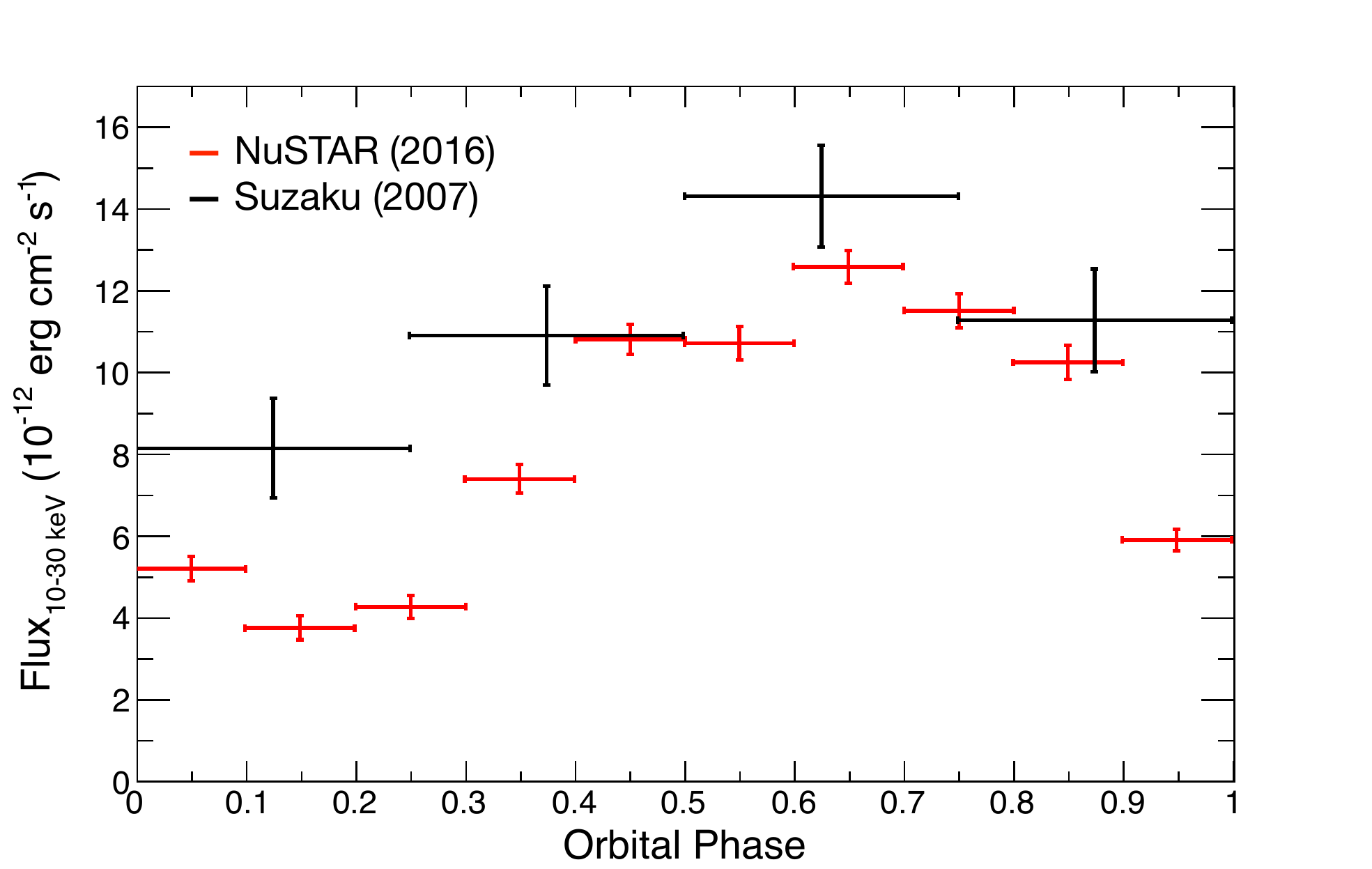}
\includegraphics[width = 8.5 cm]{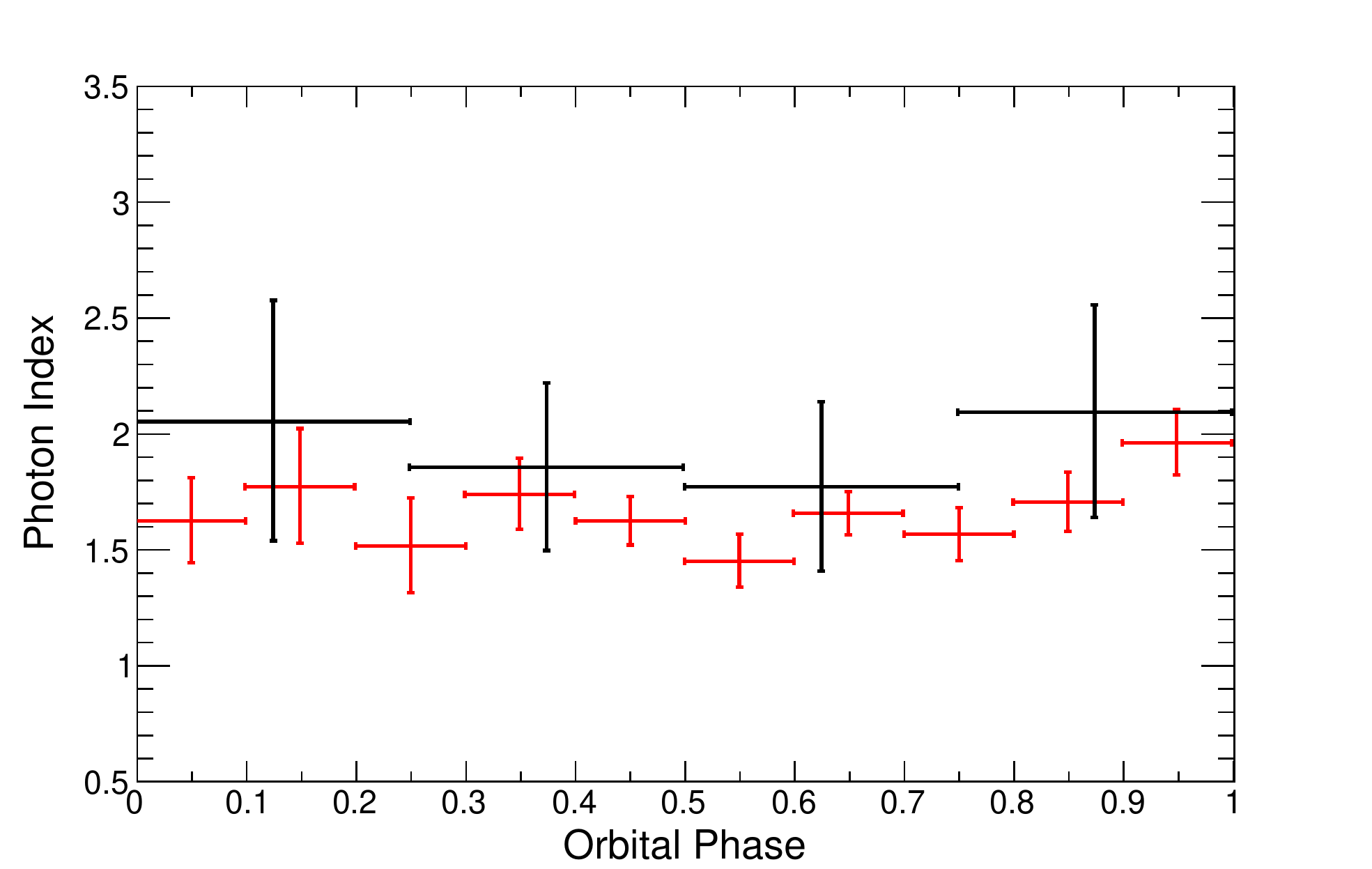}
\end{center}
\caption{The dependence of the flux (left) and the photon index (right) on the orbital phase above 10 keV.
While the red points are obtained from the 10--30 keV \nus data,
the black ones are obtained from 10--30 keV HXD/PIN data.}
\label{fig_flux_10_30_keV}
\end{figure}

Owing to the high signal-to-noise ratio of \nus,
we can investigate the flux dependence on the orbital phase above 10 keV.
We analyzed the 10--30 keV data in the same way as \S\ref{sec_comp_nustar_suzaku},
and show Figure~\ref{fig_flux_10_30_keV} which is the light curve of the flux and the photon index with a bin width of one-tenth of the orbital period.
Table~\ref{tab_spectrum_10_30_keV_nustar} describes the obtained spectral parameters.
Similar to the 3--10 keV analysis,
the flux minimum and maximum appear at $\phi =$ 0.1--0.2 and $\phi =$ 0.6--0.7, respectively.
Furthermore, the photon index does not show any significant variability.
This is again similar to the result in the 3--10 keV range.
Thus, the dependence of the emission on the orbital phase does not change between 3--10 keV and 10--30 keV.

We also analyzed the HXD/PIN data of the \suzaku observation,
and compared it to the above result.
The HXD/PIN data were screened by the standard pipeline software {\tt SUZAKU AE pipeline} 1.1.0.
The small number of events and the low signal-to-noise constrain the width of the orbital phase to 0.25 in this analysis.
As for the non X-ray background and the cosmic X-ray background,
we used {\tt LCFITDT} model provided by the \suzaku team \citep{suzaku_pinbkg} and the model defined in \citet{gruber1999}, respectively.
Since the HXD is a non-focusing detector,
the contamination of the diffuse background emission e.g. the Galactic ridge X-ray emission (GRXE) is more serious than that of \nus. 
To estimate the GRXE background,
we used the same model used in \citet{takahashistudy2009}.
In order to consider the uncertainty of the CXB and GRXE models,
we assumed that they have 10\% systematic errors.

The HXD/PIN results are shown as the black points in Figure~\ref{fig_flux_10_30_keV},
and the obtained parameters are shown in Table~\ref{tab_spectrum_10_30_keV_suzaku}.
The 10--30 keV flux of \suzaku is higher than that of \nus, by $\sim$ 25\% at the INFC.
Furthermore, the difference becomes much higher (by $\sim$69\%) at SUPC.
Although the statistical errors of the HXD/PIN result are relatively large,
the orbit-by-orbit variability in 10--30 keV might be more significant than below 10 keV (Figure~\ref{fig_flux_1_10_keV}).
Since the errors of the photon index in the HXD/PIN analysis are too large, it is difficult to discuss whether the photon index is different between the \suzaku and \nus observations.

\begin{table}[!htb]
\caption{Results of the spectral analysis of 10--30 keV \nus data.}
\label{tab_spectrum_10_30_keV_nustar}
  \begin{center}
    \begin{tabular}{c | c c c} \hline
        & \multicolumn{3}{c}{NuSTAR (10--30 keV)}\\ \hline
      Orbital Phase & Photon index & Flux (10--30 keV)$^\ast$ & $\chi^2/$dof\\ 
      INFC & $1.60\pm0.05$ & $11.31\pm{0.18}$ & $176.7/179$\\
      SUPC & $1.69\pm0.06$ & $5.75\pm{0.12}$ & $143.9/123$\\
      All Phase & $1.64\pm{0.04}$ & $8.26\pm{0.10}$ & $291.5/280$\\ \hline
      0.0--0.1& $ 1.62_{-0.18}^{+0.19} $ & $ 5.21 \pm 0.3 $           & $ 10.6/16 $ \\
      0.1--0.2& $ 1.77_{-0.24}^{+0.25} $ & $ 3.76_{-0.29}^{+0.3} $    & $  9.8/12 $ \\
      0.2--0.3& $ 1.51_{-0.20}^{+0.21} $ & $ 4.26 \pm 0.28 $          & $ 11.4/15 $ \\
      0.3--0.4& $ 1.74 \pm 0.15 $        & $ 7.38 \pm 0.35 $          & $ 22.6/26 $ \\
      0.4--0.5& $ 1.62_{-0.10}^{+0.11} $ & $ 10.79 \pm 0.37 $         & $ 55.3/45 $ \\
      0.5--0.6& $ 1.44_{-0.11}^{+0.12} $ & $ 10.71 \pm 0.41 $         & $ 43.5/36 $ \\
      0.6--0.7& $ 1.65 \pm 0.09 $        & $ 12.57 \pm 0.4 $          & $ 48.0/50 $ \\
      0.7--0.8& $ 1.57 \pm 0.11 $        & $ 11.49 \pm 0.42 $         & $ 34.8/41 $ \\
      0.8--0.9& $ 1.70 \pm 0.13 $        & $ 10.22_{-0.41}^{+0.42} $  & $ 38.4/35 $ \\
      0.9--1.0& $ 1.96 \pm 0.14 $        & $ 5.89 \pm 0.27 $          & $ 20.2/25 $ \\ \hline
    \end{tabular}
\end{center}
{\small $\ast$: Its units are $10^{-12}~\mathrm{erg~cm^{-2}~s^{-1}}$.}
\end{table}

\begin{table}[!htb]
\caption{Results of the spectral analysis of 10--30 keV \suzaku HXD/PIN data.}
\label{tab_spectrum_10_30_keV_suzaku}
  \begin{center}
    \begin{tabular}{c | c c c} \hline
        & \multicolumn{2}{c}{Suzaku HXD/PIN (10--30 keV)}\\ \hline
      Orbital Phase & Photon index & Flux (10--30 keV)$^\ast$ & $\chi^2/$dof\\ 
      INFC & $1.67\pm{0.29}$ & $14.0\pm1.2$ & $7.7/6$\\
      SUPC & $2.56_{-0.39}^{+0.40}$ & $9.7\pm1.2$ & $8.0/6$\\
      All Phase & $1.87\pm{0.21}$ & $10.7\pm1.2$ & $11.6/11$ \\ \hline
      0.0--0.25& $ 2.05_{-0.51}^{+0.52} $ & $  8.1\pm1.2 $ & $ 8.4/6 $\\
      0.25--0.5& $ 1.85\pm{0.36} $ & $ 10.9\pm1.2 $ & $ 6.1/6 $\\
      0.5--0.75& $ 1.77\pm{0.36} $ & $ 14.3\pm1.2 $ & $ 6.1/6 $\\
      0.75--1.0& $ 2.09_{-0.45}^{+0.46} $ & $ 11.3\pm1.3 $ & $ 11.5/6 $\\ \hline
    \end{tabular}
\end{center}
{\small $\ast$: Its units are $10^{-12}~\mathrm{erg~cm^{-2}~s^{-1}}$.}
\end{table}

\section{Reduction of the Contamination from a Nearby Bright Pulsar}
\label{sec_pulsar_elim}
\begin{figure}[!htbp]
    \begin{center}
    \includegraphics[height = 6.0 cm]{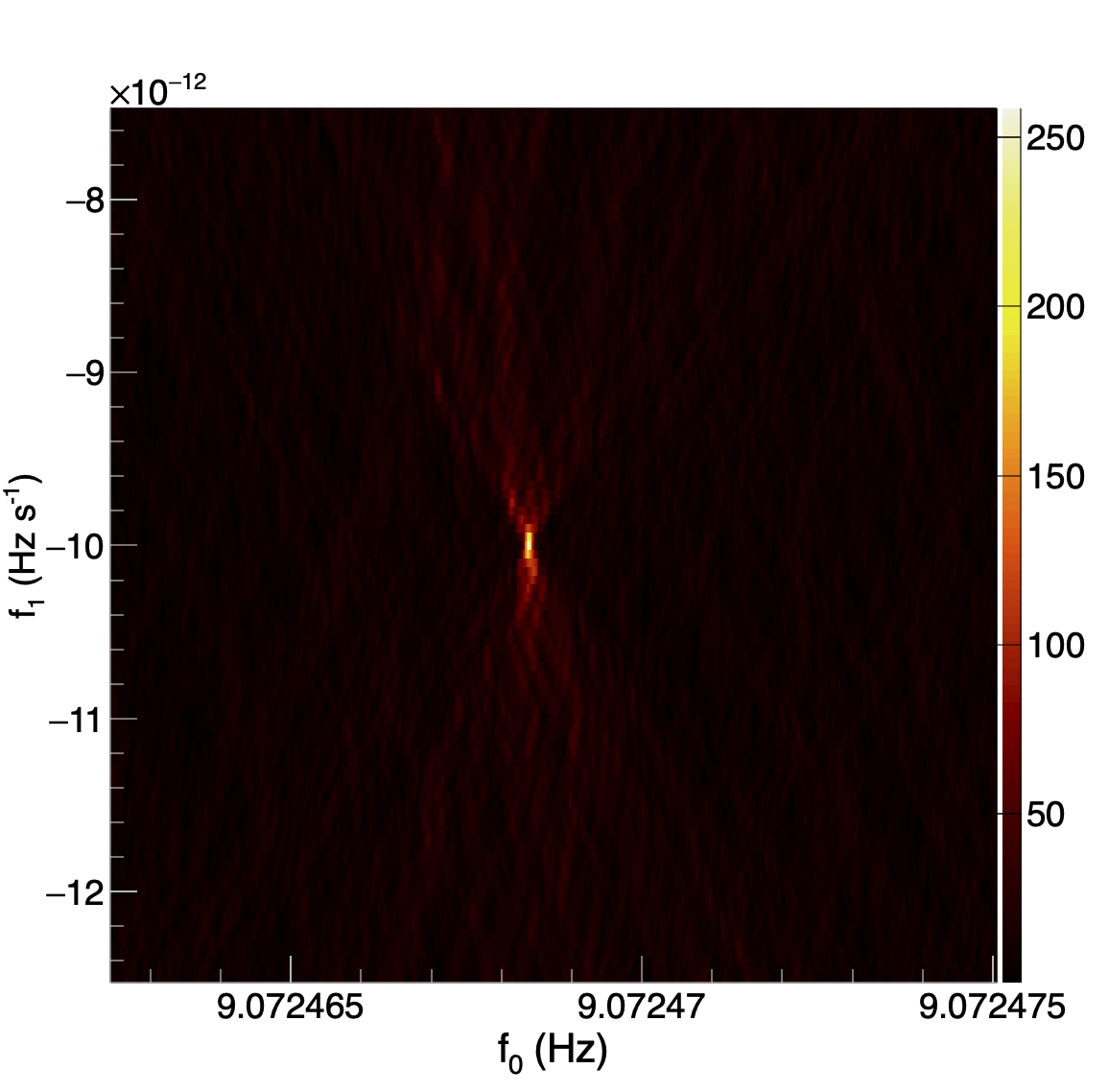}
    \includegraphics[height = 6.0 cm]{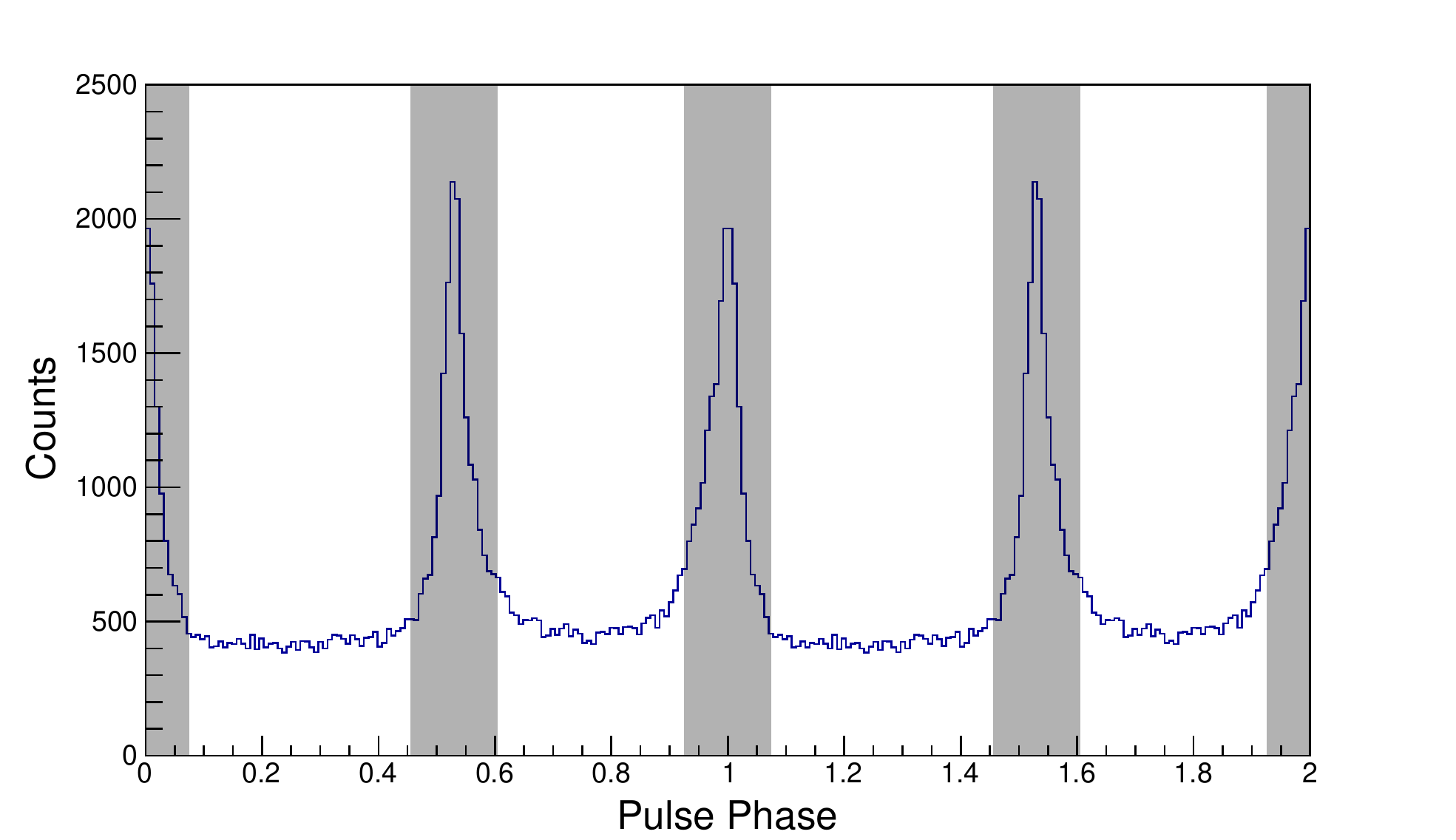}
    \end{center}
   \caption{Left: pulse model parameter search. The colors indicate $Z^2_4$ with different parameter values $f_0$ and $f_1$.
   Here $f_2$ was fixed to zero and the events in the first time interval (MET of 239557518--283456420) was used.
    Right: the pulse profile of PSR J1826-1256 in the 11 years of \fermi observation. The gray areas are the pulse phase intervals that we excluded in the analysis of \ls.}
\label{fig_profile_PSRJ1826_1256}
\end{figure}

Since \ls is located close to the Galactic plane,
it is surrounded by many gamma-ray sources.
Especially, 
there is the bright GeV pulsar PSR J1826-1256, apart from \ls by $\sim$ 2 degrees.
At 200 MeV, the flux of this pulsar is comparable to that of \ls.
Considering that the point spread function of \fermi LAT is $\sim$ 3 degrees at 200 MeV,
the gamma rays detected at the position of \ls include those emitted from PSR J1826-1256 considerably.
Thus, it is important to reduce the contamination from this pulsar \citep{fermi_lat_collaboration_2009, Hadasch2012}.
A pulse model for PSR J1826-1256 was derived by \citet{Ray_2011}, but it is valid only from 2008 August to 2010 January.
Since we need a pulse model for the 11 years of \fermi observation,
we calculated it using the LAT data.
We used the 0.2--1000 GeV events which were detected in a circular region with a radius of 0.5 degrees centered at the location of PSR J1826-1256, (RA, DEC) = (18:26:08.53,-12:56:33.0). 
The pulse profile was modeled by three parameters;
the frequency ($f_0$) and the first and second derivative of the frequency ($f_1$ and $f_2$).
In the parameter estimation,
we used the Z$^2$ statistics \citep{deJager1989} defined as
\begin{eqnarray}
\label{eq_Z2_def}
&Z^2_4 = \frac{2}{N} \displaystyle \sum_{m=1}^4 \left( \left(\displaystyle \sum_{i=0}^N \cos 2\pi m \phi_i \right)^2 + \left(\displaystyle \sum_{i=0}^N \sin 2\pi m \phi_i \right)^2 \right)\\
&\phi_i = f_0 (t_i - t_0) + \frac{1}{2} f_1 (t_i - t_0)^2 + \frac{1}{6} f_2 (t_i - t_0)^3~,
\end{eqnarray}
where $t_i$ is a time of each event and $N$ is the total event number and $t_0$ is a reference time.
\citet{Ray_2011} reported that the pulse profile has two peaks,
and in order to model it precisely we considered up to 4th harmonics in Eq.~\ref{eq_Z2_def}. 
Since this pulse model loses its accuracy after about three years,
we divided the data into subsets and obtained the parameters for every two years as follows.
First by exploring the parameter space of $f_0$, $f_1$ and fixing $f_2$ to zero,
we find tentative best-fit parameters that yield the maximum value of $Z^2_{4}$.
As an example, the left panel of Figure~\ref{fig_profile_PSRJ1826_1256} shows the obtained $Z^2_{4}$ map using the first subset.
Next, we also allow $f_2$ to vary and sweep the parameter space around the obtained values with finer steps,
and then we optimize the best-fit parameters.
The results are shown in Table~\ref{tab_par_PSRJ1826_1256}.
As shown in the right panel of Figure~\ref{fig_profile_PSRJ1826_1256},
the obtained pulse profile has two sharp peaks, which is consistent with \citet{Ray_2011}.

\begin{table*}[!htbp]
\begin{center}
\caption{The obtained parameters of the pulse timing model of PSR J1826-1256.}
\label{tab_par_PSRJ1826_1256}
\begin{tabular}{cccccc}
Mission Elapsed Time (s)$^\ast$ & $t_0$ (s) & $f_0$ ($\mathrm{s^{-1}}$) & $f_1$ ($\times 10^{-12} \mathrm{~s^{-2}}$) & $f_2$ ($\times 10^{-22} \mathrm{~s^{-3}}$) & pulse phase at $t_0$ \\ \hline
239557518--283456420 & 260314702.960 & $9.072468391(1)$ &  $-9.9967(2)$ & $1.8(2)$ & 0.0582 \\
283456420--327355322 & 305989728.367 & $9.072012021(2)$ &  $-9.9850(2)$ & $3.6(2)$ & 0.1518 \\
327355322--371254224 & 349374105.666 & $9.071579147(2)$ &  $-9.9709(2)$ & $3.0(2)$ & 0.9330 \\
371254224--415153126 & 398186297.308 & $9.071118379(1)$ & $-10.0086(3)$ & $5.4(2)$ & 0.0578 \\
415153126--459052028 & 434432794.768 & $9.070755858(1)$ &  $-9.9965(2)$ & $2.5(2)$ & 0.0505 \\
459052028--502950930 & 483220667.888 & $9.070268464(2)$ &  $-9.9826(2)$ & $3.0(2)$ & 0.4105 \\
502950930--546849832 & 523881039.723 & $9.069862816(1)$ &  $-9.9708(2)$ & $2.6(2)$ & 0.5509 \\
546849832--590748734 & 570655883.496 & $9.069406809(1)$ & $-10.0025(2)$ & $8.0(2)$ & 0.0746 \\
\hline
\end{tabular}
\end{center}
{\small $\ast$: 
Mission Elapsed Time (MET) is defined as the number of seconds since the reference time of a Modified Julian Date of 51910 in the UTC system.
}
\end{table*}

\bibliographystyle{aasjournal}

\begin{thebibliography}{}
\expandafter\ifx\csname natexlab\endcsname\relax\def\natexlab#1{#1}\fi
\providecommand{\url}[1]{\href{#1}{#1}}
\providecommand{\dodoi}[1]{doi:~\href{http://doi.org/#1}{\nolinkurl{#1}}}
\providecommand{\doeprint}[1]{\href{http://ascl.net/#1}{\nolinkurl{http://ascl.net/#1}}}
\providecommand{\doarXiv}[1]{\href{https://arxiv.org/abs/#1}{\nolinkurl{https://arxiv.org/abs/#1}}}

\bibitem[{Abdo {et~al.}(2009)Abdo, Ackermann, Ajello, Atwood, Axelsson,
  Baldini, Ballet, Barbiellini, Bastieri, Baughman, Bechtol, Bellazzini,
  Berenji, Blandford, Bloom, Bonamente, Borgland, Bregeon, Brez, Brigida,
  Bruel, Burnett, Buson, Caliandro, Cameron, Caraveo, Casandjian, Cavazzuti,
  Cecchi, {\c{C}}elik, Chaty, Chekhtman, Cheung, Chiang, Ciprini, Claus,
  Cohen-Tanugi, Cominsky, Conrad, Corbel, Corbet, Cutini, Dermer, de~Angelis,
  de~Palma, Digel, do~Couto~e Silva, Drell, Dubois, Dubus, Dumora, Farnier,
  Favuzzi, Fegan, Focke, Fortin, Frailis, Fukazawa, Funk, Fusco, Gargano,
  Gasparrini, Gehrels, Germani, Giebels, Giglietto, Giordano, Glanzman,
  Godfrey, Grenier, Grondin, Grove, Guillemot, Guiriec, Hanabata, Harding,
  Hayashida, Hays, Hill, Horan, Hughes, Jackson, J{\'{o}}hannesson, Johnson,
  Johnson, Johnson, Kamae, Katagiri, Kataoka, Kawai, Kerr, Knödlseder, Kocian,
  Kuehn, Kuss, Lande, Larsson, Latronico, Lemoine-Goumard, Longo, Loparco,
  Lott, Lovellette, Lubrano, Madejski, Makeev, Marelli, Mazziotta, McEnery,
  Meurer, Michelson, Mitthumsiri, Mizuno, Moiseev, Monte, Monzani, Morselli,
  Moskalenko, Murgia, Nolan, Norris, Nuss, Ohsugi, Omodei, Orlando, Ormes,
  Ozaki, Paneque, Panetta, Parent, Pelassa, Pepe, Pesce-Rollins, Piron, Porter,
  Rain{\`{o}}, Rando, Ray, Razzano, Rea, Reimer, Reimer, Reposeur, Ritz,
  Rochester, Rodriguez, Romani, Roth, Ryde, Sadrozinski, Sanchez, Sander,
  Parkinson, Scargle, Sgr{\`{o}}, Sierpowska-Bartosik, Siskind, Smith, Smith,
  Spandre, Spinelli, Strickman, Suson, Tajima, Takahashi, Takahashi, Tanaka,
  Tanaka, Thayer, Thompson, Tibaldo, Torres, Tosti, Tramacere, Uchiyama, Usher,
  Vasileiou, Venter, Vilchez, Vitale, Waite, Wallace, Wang, Winer, Wood,
  Ylinen, \& Ziegler}]{fermi_lat_collaboration_2009}
Abdo, A.~A., Ackermann, M., Ajello, M., {et~al.} 2009, \apj, 706, L56,
  \dodoi{10.1088/0004-637x/706/1/l56}

\bibitem[{{Abdo} {et~al.}(2011){Abdo}, {Ackermann}, {Ajello}, {Allafort},
  {Ballet}, {Barbiellini}, {Bastieri}, {Bechtol}, {Bellazzini}, {Berenji},
  {Bland ford}, {Bonamente}, {Borgland}, {Bregeon}, {Brigida}, {Bruel},
  {Buehler}, {Buson}, {Caliandro}, {Cameron}, {Camilo}, {Caraveo}, {Cecchi},
  {Charles}, {Chaty}, {Chekhtman}, {Chernyakova}, {Cheung}, {Chiang},
  {Ciprini}, {Claus}, {Cohen-Tanugi}, {Cominsky}, {Corbel}, {Cutini},
  {D'Ammando}, {de Angelis}, {den Hartog}, {de Palma}, {Dermer}, {Digel},
  {Silva}, {Dormody}, {Drell}, {Drlica-Wagner}, {Dubois}, {Dubus}, {Dumora},
  {Enoto}, {Espinoza}, {Favuzzi}, {Fegan}, {Ferrara}, {Focke}, {Fortin},
  {Fukazawa}, {Funk}, {Fusco}, {Gargano}, {Gasparrini}, {Gehrels}, {Germani},
  {Giglietto}, {Giommi}, {Giordano}, {Giroletti}, {Glanzman}, {Godfrey},
  {Grenier}, {Grondin}, {Grove}, {Grundstrom}, {Guiriec}, {Gwon}, {Hadasch},
  {Harding}, {Hayashida}, {Hays}, {J{\'o}hannesson}, {Johnson}, {Johnson},
  {Johnston}, {Kamae}, {Katagiri}, {Kataoka}, {Keith}, {Kerr},
  {Kn{\"o}dlseder}, {Kramer}, {Kuss}, {Lande}, {Lee}, {Lemoine-Goumard},
  {Longo}, {Loparco}, {Lovellette}, {Lubrano}, {Manchester}, {Marelli},
  {Mazziotta}, {Michelson}, {Mitthumsiri}, {Mizuno}, {Moiseev}, {Monte},
  {Monzani}, {Morselli}, {Moskalenko}, {Murgia}, {Nakamori}, {Naumann-Godo},
  {Neronov}, {Nolan}, {Norris}, {Noutsos}, {Nuss}, {Ohsugi}, {Okumura},
  {Omodei}, {Orlando}, {Paneque}, {Parent}, {Pesce-Rollins}, {Pierbattista},
  {Piron}, {Porter}, {Possenti}, {Rain{\`o}}, {Rando}, {Ray}, {Razzano},
  {Razzaque}, {Reimer}, {Reimer}, {Reposeur}, {Ritz}, {Sadrozinski}, {Scargle},
  {Sgr{\`o}}, {Shannon}, {Siskind}, {Smith}, {Spandre}, {Spinelli},
  {Strickman}, {Suson}, {Takahashi}, {Tanaka}, {Thayer}, {Thayer}, {Thompson},
  {Thorsett}, {Tibaldo}, {Tibolla}, {Torres}, {Tosti}, {Troja}, {Uchiyama},
  {Usher}, {Vand enbroucke}, {Vasileiou}, {Vianello}, {Vitale}, {Waite},
  {Wang}, {Winer}, {Wolff}, {Wood}, {Wood}, {Yang}, {Ziegler}, \&
  {Zimmer}}]{2011ApJ...736L..11A}
{Abdo}, A.~A., {Ackermann}, M., {Ajello}, M., {et~al.} 2011, \apjl, 736, L11,
  \dodoi{10.1088/2041-8205/736/1/L11}

\bibitem[{{Abdo} {et~al.}(2013){Abdo}, {Ajello}, {Allafort}, {Baldini},
  {Ballet}, {Barbiellini}, {Baring}, {Bastieri}, {Belfiore}, {Bellazzini},
  {Bhattacharyya}, {Bissaldi}, {Bloom}, {Bonamente}, {Bottacini}, {Brandt},
  {Bregeon}, {Brigida}, {Bruel}, {Buehler}, {Burgay}, {Burnett}, {Busetto},
  {Buson}, {Caliandro}, {Cameron}, {Camilo}, {Caraveo}, {Casandjian}, {Cecchi},
  {{\c{C}}elik}, {Charles}, {Chaty}, {Chaves}, {Chekhtman}, {Chen}, {Chiang},
  {Chiaro}, {Ciprini}, {Claus}, {Cognard}, {Cohen-Tanugi}, {Cominsky},
  {Conrad}, {Cutini}, {D'Ammando}, {de Angelis}, {DeCesar}, {De Luca}, {den
  Hartog}, {de Palma}, {Dermer}, {Desvignes}, {Digel}, {Di Venere}, {Drell},
  {Drlica-Wagner}, {Dubois}, {Dumora}, {Espinoza}, {Falletti}, {Favuzzi},
  {Ferrara}, {Focke}, {Franckowiak}, {Freire}, {Funk}, {Fusco}, {Gargano},
  {Gasparrini}, {Germani}, {Giglietto}, {Giommi}, {Giordano}, {Giroletti},
  {Glanzman}, {Godfrey}, {Gotthelf}, {Grenier}, {Grondin}, {Grove},
  {Guillemot}, {Guiriec}, {Hadasch}, {Hanabata}, {Harding}, {Hayashida},
  {Hays}, {Hessels}, {Hewitt}, {Hill}, {Horan}, {Hou}, {Hughes}, {Jackson},
  {Janssen}, {Jogler}, {J{\'o}hannesson}, {Johnson}, {Johnson}, {Johnson},
  {Johnson}, {Johnston}, {Kamae}, {Kataoka}, {Keith}, {Kerr}, {Kn{\"o}dlseder},
  {Kramer}, {Kuss}, {Lande}, {Larsson}, {Latronico}, {Lemoine-Goumard},
  {Longo}, {Loparco}, {Lovellette}, {Lubrano}, {Lyne}, {Manchester}, {Marelli},
  {Massaro}, {Mayer}, {Mazziotta}, {McEnery}, {McLaughlin}, {Mehault},
  {Michelson}, {Mignani}, {Mitthumsiri}, {Mizuno}, {Moiseev}, {Monzani},
  {Morselli}, {Moskalenko}, {Murgia}, {Nakamori}, {Nemmen}, {Nuss}, {Ohno},
  {Ohsugi}, {Orienti}, {Orlando}, {Ormes}, {Paneque}, {Panetta}, {Parent},
  {Perkins}, {Pesce-Rollins}, {Pierbattista}, {Piron}, {Pivato}, {Pletsch},
  {Porter}, {Possenti}, {Rain{\`o}}, {Rando}, {Ransom}, {Ray}, {Razzano},
  {Rea}, {Reimer}, {Reimer}, {Renault}, {Reposeur}, {Ritz}, {Romani}, {Roth},
  {Rousseau}, {Roy}, {Ruan}, {Sartori}, {Saz Parkinson}, {Scargle}, {Schulz},
  {Sgr{\`o}}, {Shannon}, {Siskind}, {Smith}, {Spandre}, {Spinelli}, {Stappers},
  {Strong}, {Suson}, {Takahashi}, {Thayer}, {Thayer}, {Theureau}, {Thompson},
  {Thorsett}, {Tibaldo}, {Tibolla}, {Tinivella}, {Torres}, {Tosti}, {Troja},
  {Uchiyama}, {Usher}, {Vandenbroucke}, {Vasileiou}, {Venter}, {Vianello},
  {Vitale}, {Wang}, {Weltevrede}, {Winer}, {Wolff}, {Wood}, {Wood}, {Wood}, \&
  {Yang}}]{2PC2013}
{Abdo}, A.~A., {Ajello}, M., {Allafort}, A., {et~al.} 2013, \apjs, 208, 17,
  \dodoi{10.1088/0067-0049/208/2/17}

\bibitem[{{Abdo} {et~al.}(2015){Abdo}, {Ackermann}, {Ajello}, {Allafort},
  {Amin}, {Baldini}, {Barbiellini}, {Bastieri}, {Bechtol}, {Bellazzini}, {Bland
  ford}, {Bonamente}, {Borgland}, {Bregeon}, {Brigida}, {Buehler}, {Bulmash},
  {Buson}, {Caliandro}, {Cameron}, {Caraveo}, {Cavazzuti}, {Cecchi}, {Charles},
  {Cheung}, {Chiang}, {Chiaro}, {Ciprini}, {Claus}, {Cohen-Tanugi}, {Conrad},
  {Corbet}, {Cutini}, {D'Ammando}, {de Angelis}, {de Palma}, {Dermer}, {Drell},
  {Drlica-Wagner}, {Favuzzi}, {Finke}, {Focke}, {Fukazawa}, {Fusco}, {Gargano},
  {Gasparrini}, {Gehrels}, {Giglietto}, {Giordano}, {Giroletti}, {Glanzman},
  {Grenier}, {Grove}, {Guiriec}, {Hadasch}, {Hayashida}, {Hays}, {Hughes},
  {Inoue}, {Jackson}, {Jogler}, {J{\'o}hannesson}, {Johnson}, {Kamae},
  {Kn{\"o}dlseder}, {Kuss}, {Lande}, {Larsson}, {Latronico}, {Longo},
  {Loparco}, {Lott}, {Lovellette}, {Lubrano}, {Madejski}, {Mazziotta},
  {Mehault}, {Michelson}, {Mizuno}, {Monzani}, {Morselli}, {Moskalenko},
  {Murgia}, {Nemmen}, {Nuss}, {Ohno}, {Ohsugi}, {Paneque}, {Perkins},
  {Pesce-Rollins}, {Piron}, {Pivato}, {Porter}, {Rain{\`o}}, {Rando},
  {Razzano}, {Reimer}, {Reimer}, {Reyes}, {Ritz}, {Romoli}, {Roth}, {Saz
  Parkinson}, {Sgr{\`o}}, {Siskind}, {Spandre}, {Spinelli}, {Takahashi},
  {Takeuchi}, {Tanaka}, {Thayer}, {Thayer}, {Thompson}, {Tibaldo}, {Tinivella},
  {Torres}, {Tosti}, {Troja}, {Tronconi}, {Usher}, {Vand enbroucke},
  {Vasileiou}, {Vianello}, {Vitale}, {Waite}, {Werner}, {Winer}, \&
  {Wood}}]{PKS2015}
{Abdo}, A.~A., {Ackermann}, M., {Ajello}, M., {et~al.} 2015, \apj, 799, 143,
  \dodoi{10.1088/0004-637X/799/2/143}

\bibitem[{Abdollahi {et~al.}(2020)Abdollahi, Acero, Ackermann, Ajello, Atwood,
  Axelsson, Baldini, Ballet, Barbiellini, Bastieri, Gonzalez, Bellazzini,
  Berretta, Bissaldi, Blandford, Bloom, Bonino, Bottacini, Brandt, Bregeon,
  Bruel, Buehler, Burnett, Buson, Cameron, Caputo, Caraveo, Casandjian, Castro,
  Cavazzuti, Charles, Chaty, Chen, Cheung, Chiaro, Ciprini, Cohen-Tanugi,
  Cominsky, Coronado-Bl{\'{a}}zquez, Costantin, Cuoco, Cutini, D'Ammando,
  DeKlotz, de~la Torre~Luque, de~Palma, Desai, Digel, Lalla, Mauro, Venere,
  Dom{\'{\i}}nguez, Dumora, Dirirsa, Fegan, Ferrara, Franckowiak, Fukazawa,
  Funk, Fusco, Gargano, Gasparrini, Giglietto, Giommi, Giordano, Giroletti,
  Glanzman, Green, Grenier, Griffin, Grondin, Grove, Guiriec, Harding, Hayashi,
  Hays, Hewitt, Horan, J{\'{o}}hannesson, Johnson, Kamae, Kerr, Kocevski,
  Kovac'evic', Kuss, Landriu, Larsson, Latronico, Lemoine-Goumard, Li,
  Liodakis, Longo, Loparco, Lott, Lovellette, Lubrano, Madejski, Maldera,
  Malyshev, Manfreda, Marchesini, Marcotulli, Mart{\'{\i}}-Devesa, Martin,
  Massaro, Mazziotta, McEnery, Mereu, Meyer, Michelson, Mirabal, Mizuno,
  Monzani, Morselli, Moskalenko, Negro, Nuss, Ojha, Omodei, Orienti, Orlando,
  Ormes, Palatiello, Paliya, Paneque, Pei, Pe{\~{n}}a-Herazo, Perkins, Persic,
  Pesce-Rollins, Petrosian, Petrov, Piron, Poon, Porter, Principe, Rain{\`{o}},
  Rando, Razzano, Razzaque, Reimer, Reimer, Remy, Reposeur, Romani, Parkinson,
  Schinzel, Serini, Sgr{\`{o}}, Siskind, Smith, Spandre, Spinelli, Strong,
  Suson, Tajima, Takahashi, Tak, Thayer, Thompson, Tibaldo, Torres, Torresi,
  Valverde, Klaveren, van Zyl, Wood, Yassine, \& Zaharijas}]{fermi4thcatalog}
Abdollahi, S., Acero, F., Ackermann, M., {et~al.} 2020, \apjs, 247, 33,
  \dodoi{10.3847/1538-4365/ab6bcb}

\bibitem[{{Acciari} {et~al.}(2008){Acciari}, {Beilicke}, {Blaylock},
  {Bradbury}, {Buckley}, {Bugaev}, {Butt}, {Byrum}, {Celik}, {Cesarini},
  {Ciupik}, {Chow}, {Cogan}, {Colin}, {Cui}, {Daniel}, {Duke}, {Ergin},
  {Falcone}, {Fegan}, {Finley}, {Fortin}, {Fortson}, {Gall}, {Gibbs}, {Gilland
  ers}, {Grube}, {Guenette}, {Hanna}, {Hays}, {Holder}, {Horan}, {Hughes},
  {Hui}, {Humensky}, {Kaaret}, {Kieda}, {Kildea}, {Konopelko}, {Krawczynski},
  {Krennrich}, {Lang}, {LeBohec}, {Lee}, {Maier}, {McCann}, {McCutcheon},
  {Millis}, {Moriarty}, {Mukherjee}, {Nagai}, {Ong}, {Pandel}, {Perkins},
  {Pizlo}, {Pohl}, {Quinn}, {Ragan}, {Reynolds}, {Rose}, {Schroedter},
  {Sembroski}, {Smith}, {Steele}, {Swordy}, {Toner}, {Valcarcel}, {Vassiliev},
  {Wagner}, {Wakely}, {Ward}, {Weekes}, {Weinstein}, {White}, {Williams},
  {Wissel}, {Wood}, \& {Zitzer}}]{2008ApJ...679.1427A}
{Acciari}, V.~A., {Beilicke}, M., {Blaylock}, G., {et~al.} 2008, \apj, 679,
  1427, \dodoi{10.1086/587736}

\bibitem[{{Aharonian} {et~al.}(2005{\natexlab{a}}){Aharonian}, {Akhperjanian},
  {Aye}, {Bazer-Bachi}, {Beilicke}, {Benbow}, {Berge}, {Berghaus},
  {Bernl{\"o}hr}, {Boisson}, {Bolz}, {Borrel}, {Braun}, {Breitling}, {Brown},
  {Gordo}, {Chadwick}, {Chounet}, {Cornils}, {Costamante}, {Degrange},
  {Dickinson}, {Djannati-Ata{\"\i}}, {Drury}, {Dubus}, {Emmanoulopoulos},
  {Espigat}, {Feinstein}, {Fleury}, {Fontaine}, {Fuchs}, {Funk}, {Gallant},
  {Giebels}, {Gillessen}, {Glicenstein}, {Goret}, {Hadjichristidis}, {Hauser},
  {Heinzelmann}, {Henri}, {Hermann}, {Hinton}, {Hofmann}, {Holleran}, {Horns},
  {Jacholkowska}, {de Jager}, {Kh{\'e}lifi}, {Komin}, {Konopelko}, {Latham},
  {Le Gallou}, {Lemi{\`e}re}, {Lemoine-Goumard}, {Leroy}, {Lohse}, {Marcowith},
  {Martin}, {Martineau-Huynh}, {Masterson}, {McComb}, {de Naurois}, {Nolan},
  {Noutsos}, {Orford}, {Osborne}, {Ouchrif}, {Panter}, {Pelletier}, {Pita},
  {P{\"u}hlhofer}, {Punch}, {Raubenheimer}, {Raue}, {Raux}, {Rayner}, {Reimer},
  {Reimer}, {Ripken}, {Rob}, {Rolland}, {Rowell}, {Sahakian}, {Saug{\'e}},
  {Schlenker}, {Schlickeiser}, {Schuster}, {Schwanke}, {Siewert}, {Sol},
  {Spangler}, {Steenkamp}, {Stegmann}, {Tavernet}, {Terrier}, {Th{\'e}oret},
  {Tluczykont}, {Vasileiadis}, {Venter}, {Vincent}, {V{\"o}lk}, \&
  {Wagner}}]{hessls50392005}
{Aharonian}, F., {Akhperjanian}, A.~G., {Aye}, K.~M., {et~al.}
  2005{\natexlab{a}}, Science, 309, 746, \dodoi{10.1126/science.1113764}

\bibitem[{{Aharonian} {et~al.}(2005{\natexlab{b}}){Aharonian}, {Akhperjanian},
  {Aye}, {Bazer-Bachi}, {Beilicke}, {Benbow}, {Berge}, {Berghaus},
  {Bernl{\"o}hr}, {Boisson}, {Bolz}, {Braun}, {Breitling}, {Brown}, {Bussons
  Gordo}, {Chadwick}, {Chounet}, {Cornils}, {Costamante}, {Degrange},
  {Djannati-Ata{\"\i}}, {O'C. Drury}, {Dubus}, {Emmanoulopoulos}, {Espigat},
  {Feinstein}, {Fleury}, {Fontaine}, {Fuchs}, {Funk}, {Gallant}, {Giebels},
  {Gillessen}, {Glicenstein}, {Goret}, {Hadjichristidis}, {Hauser},
  {Heinzelmann}, {Henri}, {Hermann}, {Hinton}, {Hofmann}, {Holleran}, {Horns},
  {de Jager}, {Johnston}, {Kh{\'e}lifi}, {Kirk}, {Komin}, {Konopelko},
  {Latham}, {Le Gallou}, {Lemi{\`e}re}, {Lemoine-Goumard}, {Leroy},
  {Martineau-Huynh}, {Lohse}, {Marcowith}, {Masterson}, {McComb}, {de Naurois},
  {Nolan}, {Noutsos}, {Orford}, {Osborne}, {Ouchrif}, {Panter}, {Pelletier},
  {Pita}, {P{\"u}hlhofer}, {Punch}, {Raubenheimer}, {Raue}, {Raux}, {Rayner},
  {Redondo}, {Reimer}, {Reimer}, {Ripken}, {Rob}, {Rolland}, {Rowell},
  {Sahakian}, {Saug{\'e}}, {Schlenker}, {Schlickeiser}, {Schuster}, {Schwanke},
  {Siewert}, {Skj{\ae}raasen}, {Sol}, {Steenkamp}, {Stegmann}, {Tavernet},
  {Terrier}, {Th{\'e}oret}, {Tluczykont}, {Vasileiadis}, {Venter}, {Vincent},
  {V{\"o}lk}, \& {Wagner}}]{PSRB1259HESS2005}
---. 2005{\natexlab{b}}, \aap, 442, 1, \dodoi{10.1051/0004-6361:20052983}

\bibitem[{{Aharonian} {et~al.}(2006){Aharonian}, {Akhperjanian}, {Bazer-Bachi},
  {Beilicke}, {Benbow}, {Berge}, {Bernl{\"o}hr}, {Boisson}, {Bolz}, {Borrel},
  {Braun}, {Brown}, {B{\"u}hler}, {B{\"u}sching}, {Carrigan}, {Chadwick},
  {Chounet}, {Cornils}, {Costamante}, {Degrange}, {Dickinson},
  {Djannati-Ata{\"\i}}, {O'C. Drury}, {Dubus}, {Egberts}, {Emmanoulopoulos},
  {Espigat}, {Feinstein}, {Ferrero}, {Fiasson}, {Fontaine}, {Funk}, {Funk},
  {F{\"u}{\ss}ling}, {Gallant}, {Giebels}, {Glicenstein}, {Goret},
  {Hadjichristidis}, {Hauser}, {Hauser}, {Heinzelmann}, {Henri}, {Hermann},
  {Hinton}, {Hoffmann}, {Hofmann}, {Holleran}, {Horns}, {Jacholkowska}, {de
  Jager}, {Kendziorra}, {Kh{\'e}lifi}, {Komin}, {Konopelko}, {Kosack},
  {Latham}, {Le Gallou}, {Lemi{\`e}re}, {Lemoine-Goumard}, {Lohse}, {Martin},
  {Martineau-Huynh}, {Marcowith}, {Masterson}, {Maurin}, {McComb}, {Moulin},
  {de Naurois}, {Nedbal}, {Nolan}, {Noutsos}, {Orford}, {Osborne}, {Ouchrif},
  {Panter}, {Pelletier}, {Pita}, {P{\"u}hlhofer}, {Punch}, {Raubenheimer},
  {Raue}, {Rayner}, {Reimer}, {Reimer}, {Ripken}, {Rob}, {Rolland}, {Rowell},
  {Sahakian}, {Santangelo}, {Saug{\'e}}, {Schlenker}, {Schlickeiser},
  {Schr{\"o}der}, {Schwanke}, {Schwarzburg}, {Shalchi}, {Sol}, {Spangler},
  {Spanier}, {Steenkamp}, {Stegmann}, {Superina}, {Tavernet}, {Terrier},
  {Tluczykont}, {van Eldik}, {Vasileiadis}, {Venter}, {Vincent}, {V{\"o}lk},
  {Wagner}, \& {Ward}}]{hess2006}
{Aharonian}, F., {Akhperjanian}, A.~G., {Bazer-Bachi}, A.~R., {et~al.} 2006,
  \aap, 460, 743, \dodoi{10.1051/0004-6361:20065940}

\bibitem[{Aharonian(2004)}]{felixblackbook}
Aharonian, F.~A. 2004, Very High Energy Cosmic Gamma Radiation (WORLD
  SCIENTIFIC), \dodoi{10.1142/4657}

\bibitem[{Aharonian {et~al.}(2002)Aharonian, Belyanin, Derishev, Kocharovsky,
  \& Kocharovsky}]{Aharonian2002}
Aharonian, F.~A., Belyanin, A.~A., Derishev, E.~V., Kocharovsky, V.~V., \&
  Kocharovsky, V.~V. 2002, Phys. Rev. D, 66, 023005,
  \dodoi{10.1103/PhysRevD.66.023005}

\bibitem[{{Albert} {et~al.}(2006){Albert}, {Aliu}, {Anderhub}, {Antoranz},
  {Armada}, {Asensio}, {Baixeras}, {Barrio}, {Bartelt}, {Bartko}, {Bastieri},
  {Bavikadi}, {Bednarek}, {Berger}, {Bigongiari}, {Biland}, {Bisesi}, {Bock},
  {Bordas}, {Bosch-Ramon}, {Bretz}, {Britvitch}, {Camara}, {Carmona},
  {Chilingarian}, {Ciprini}, {Coarasa}, {Commichau}, {Contreras}, {Cortina},
  {Curtef}, {Danielyan}, {Dazzi}, {De Angelis}, {de los Reyes}, {De Lotto},
  {Domingo-Santamar{\'\i}a}, {Dorner}, {Doro}, {Errand o}, {Fagiolini},
  {Ferenc}, {Fern{\'a}ndez}, {Firpo}, {Flix}, {Fonseca}, {Font}, {Fuchs},
  {Galante}, {Garczarczyk}, {Gaug}, {Giller}, {Goebel}, {Hakobyan},
  {Hayashida}, {Hengstebeck}, {H{\"o}hne}, {Hose}, {Hsu}, {Isar}, {Jacon},
  {Kalekin}, {Kosyra}, {Kranich}, {Laatiaoui}, {Laille}, {Lenisa}, {Liebing},
  {Lindfors}, {Lombardi}, {Longo}, {L{\'o}pez}, {L{\'o}pez}, {Lorenz},
  {Lucarelli}, {Majumdar}, {Maneva}, {Mannheim}, {Mansutti}, {Mariotti},
  {Mart{\'\i}nez}, {Mase}, {Mazin}, {Merck}, {Meucci}, {Meyer}, {Miranda},
  {Mirzoyan}, {Mizobuchi}, {Moralejo}, {Nilsson}, {O{\~n}a-Wilhelmi},
  {Ordu{\~n}a}, {Otte}, {Oya}, {Paneque}, {Paoletti}, {Paredes}, {Pasanen},
  {Pascoli}, {Pauss}, {Pavel}, {Pegna}, {Persic}, {Peruzzo}, {Piccioli},
  {Poller}, {Pooley}, {Prandini}, {Raymers}, {Rhode}, {Rib{\'o}}, {Rico},
  {Riegel}, {Rissi}, {Robert}, {Romero}, {R{\"u}gamer}, {Saggion},
  {S{\'a}nchez}, {Sartori}, {Scalzotto}, {Scapin}, {Schmitt}, {Schweizer},
  {Shayduk}, {Shinozaki}, {Shore}, {Sidro}, {Sillanp{\"a}{\"a}}, {Sobczynska},
  {Stamerra}, {Stark}, {Takalo}, {Temnikov}, {Tescaro}, {Teshima}, {Tonello},
  {Torres}, {Torres}, {Turini}, {Vankov}, {Vitale}, {Wagner}, {Wibig},
  {Wittek}, {Zanin}, \& {Zapatero}}]{2006Sci...312.1771A}
{Albert}, J., {Aliu}, E., {Anderhub}, H., {et~al.} 2006, Science, 312, 1771,
  \dodoi{10.1126/science.1128177}

\bibitem[{Aragona {et~al.}(2009)Aragona, McSwain, Grundstrom, Marsh,
  Roettenbacher, Hessler, Boyajian, \& Ray}]{aragonaorbits2009}
Aragona, C., McSwain, M.~V., Grundstrom, E.~D., {et~al.} 2009, \apj, 698, 514,
  \dodoi{10.1088/0004-637x/698/1/514}

\bibitem[{{Arons}(2012)}]{2012SSRv..173..341A}
{Arons}, J. 2012, \ssr, 173, 341, \dodoi{10.1007/s11214-012-9885-1}

\bibitem[{{Atwood} {et~al.}(2013){Atwood}, {Albert}, {Baldini}, {Tinivella},
  {Bregeon}, {Pesce-Rollins}, {Sgr{\`o}}, {Bruel}, {Charles}, {Drlica-Wagner},
  {Franckowiak}, {Jogler}, {Rochester}, {Usher}, {Wood}, {Cohen-Tanugi}, \&
  {Zimmer}}]{pass8_1}
{Atwood}, W., {Albert}, A., {Baldini}, L., {et~al.} 2013, arXiv e-prints,
  arXiv:1303.3514.
\newblock \doarXiv{1303.3514}

\bibitem[{Becker \& Wolff(2007)}]{Becker_2007}
Becker, P.~A., \& Wolff, M.~T. 2007, \apj, 654, 435, \dodoi{10.1086/509108}

\bibitem[{{Bosch-Ramon} \& {Khangulyan}(2009)}]{boschramon2009}
{Bosch-Ramon}, V., \& {Khangulyan}, D. 2009, International Journal of Modern
  Physics D, 18, 347, \dodoi{10.1142/S0218271809014601}

\bibitem[{{Bruel} {et~al.}(2018){Bruel}, {Burnett}, {Digel}, {Johannesson},
  {Omodei}, \& {Wood}}]{pass8_2}
{Bruel}, P., {Burnett}, T.~H., {Digel}, S.~W., {et~al.} 2018, arXiv e-prints,
  arXiv:1810.11394.
\newblock \doarXiv{1810.11394}

\bibitem[{{Caliandro} {et~al.}(2015){Caliandro}, {Cheung}, {Li}, {Scargle},
  {Torres}, {Wood}, \& {Chernyakova}}]{2015ApJ...811...68C}
{Caliandro}, G.~A., {Cheung}, C.~C., {Li}, J., {et~al.} 2015, \apj, 811, 68,
  \dodoi{10.1088/0004-637X/811/1/68}

\bibitem[{Casares {et~al.}(2005)Casares, Ribo, Ribas, Paredes, Marti, \&
  Herrero}]{casarespossible2005}
Casares, J., Ribo, M., Ribas, I., {et~al.} 2005, \mnras, 364, 899

\bibitem[{{Chang} {et~al.}(2016){Chang}, {Zhang}, {Ji}, {Chen}, {Kretschmar},
  {Kuulkers}, {Collmar}, \& {Liu}}]{Chang2016}
{Chang}, Z., {Zhang}, S., {Ji}, L., {et~al.} 2016, \mnras, 463, 495,
  \dodoi{10.1093/mnras/stw2009}

\bibitem[{{Chernyakova} {et~al.}(2020){Chernyakova}, {Malyshev}, {Mc Keague},
  {van Soelen}, {Marais}, {Martin-Carrillo}, \& {Murphy}}]{2020MNRAS.497..648C}
{Chernyakova}, M., {Malyshev}, D., {Mc Keague}, S., {et~al.} 2020, \mnras, 497,
  648, \dodoi{10.1093/mnras/staa1876}

\bibitem[{Collmar \& Zhang(2014)}]{collmarls2014}
Collmar, W., \& Zhang, S. 2014, A\&A, 565, A38

\bibitem[{{de Jager} {et~al.}(1989){de Jager}, {Raubenheimer}, \&
  {Swanepoel}}]{deJager1989}
{de Jager}, O.~C., {Raubenheimer}, B.~C., \& {Swanepoel}, J.~W.~H. 1989, A\&A,
  221, 180

\bibitem[{{Dubus}(2006{\natexlab{a}})}]{Dubus2006b}
{Dubus}, G. 2006{\natexlab{a}}, \aap, 456, 801,
  \dodoi{10.1051/0004-6361:20054779}

\bibitem[{{Dubus}(2006{\natexlab{b}})}]{Dubus2006}
---. 2006{\natexlab{b}}, \aap, 451, 9, \dodoi{10.1051/0004-6361:20054233}

\bibitem[{{Dubus}(2013)}]{Dubus2013}
---. 2013, \aapr, 21, 64, \dodoi{10.1007/s00159-013-0064-5}

\bibitem[{{Dubus} {et~al.}(2015){Dubus}, {Lamberts}, \& {Fromang}}]{dubus2015}
{Dubus}, G., {Lamberts}, A., \& {Fromang}, S. 2015, A\&A, 581, A27

\bibitem[{{Falanga} {et~al.}(2021){Falanga}, {Bykov}, {Li}, {Krassilchtchikov},
  {Petrov}, \& {Bozzo}}]{Falanga2021}
{Falanga}, M., {Bykov}, A.~M., {Li}, Z., {et~al.} 2021, arXiv e-prints,
  arXiv:2104.07711.
\newblock \doarXiv{2104.07711}

\bibitem[{{Fukazawa} {et~al.}(2009){Fukazawa}, {Mizuno}, {Watanabe}, {Kokubun},
  {Takahashi}, {Kawano}, {Nishino}, {Sasada}, {Shirai}, {Takahashi}, {Umeki},
  {Yamasaki}, {Yasuda}, {Bamba}, {Ohno}, {Takahashi}, {Ushio}, {Enoto},
  {Kitaguchi}, {Makishima}, {Nakazawa}, {Uehara}, {Yamada}, {Yuasa}, {Isobe},
  {Kawaharada}, {Tanaka}, {Tashiro}, {Terada}, \& {Yamaoka}}]{suzaku_pinbkg}
{Fukazawa}, Y., {Mizuno}, T., {Watanabe}, S., {et~al.} 2009, \pasj, 61, S17,
  \dodoi{10.1093/pasj/61.sp1.S17}

\bibitem[{{Gruber} {et~al.}(1999){Gruber}, {Matteson}, {Peterson}, \&
  {Jung}}]{gruber1999}
{Gruber}, D.~E., {Matteson}, J.~L., {Peterson}, L.~E., \& {Jung}, G.~V. 1999,
  \apj, 520, 124, \dodoi{10.1086/307450}

\bibitem[{{Hadasch} {et~al.}(2012){Hadasch}, {Torres}, {Tanaka}, {Corbet},
  {Hill}, {Dubois}, {Dubus}, {Glanzman}, {Corbel}, {Li}, {Chen}, {Zhang},
  {Caliandro}, {Kerr}, {Richards}, {Max-Moerbeck}, {Readhead}, \&
  {Pooley}}]{Hadasch2012}
{Hadasch}, D., {Torres}, D.~F., {Tanaka}, T., {et~al.} 2012, \apj, 749, 54,
  \dodoi{10.1088/0004-637X/749/1/54}

\bibitem[{{Johnson} {et~al.}(2018){Johnson}, {Wood}, {Kerr}, {Corbet},
  {Cheung}, {Ray}, \& {Omodei}}]{2018ApJ...863...27J}
{Johnson}, T.~J., {Wood}, K.~S., {Kerr}, M., {et~al.} 2018, \apj, 863, 27,
  \dodoi{10.3847/1538-4357/aad185}

\bibitem[{{Khangulyan} {et~al.}(2008){Khangulyan}, {Aharonian}, \&
  {Bosch-Ramon}}]{khangulyan2008}
{Khangulyan}, D., {Aharonian}, F., \& {Bosch-Ramon}, V. 2008, \mnras, 383, 467

\bibitem[{{Khangulyan} {et~al.}(2014){Khangulyan}, {Aharonian}, \&
  {Kelner}}]{khangulyan2014}
{Khangulyan}, D., {Aharonian}, F.~A., \& {Kelner}, S.~R. 2014, \apj, 783, 100,
  \dodoi{10.1088/0004-637X/783/2/100}

\bibitem[{{Khangulyan} {et~al.}(2007){Khangulyan}, {Hnatic}, {Aharonian}, \&
  {Bogovalov}}]{2007MNRAS.380..320K}
{Khangulyan}, D., {Hnatic}, S., {Aharonian}, F., \& {Bogovalov}, S. 2007,
  \mnras, 380, 320, \dodoi{10.1111/j.1365-2966.2007.12075.x}

\bibitem[{{Kirk} {et~al.}(1999){Kirk}, {Ball}, \&
  {Skj{\ae}raasen}}]{1999APh....10...31K}
{Kirk}, J.~G., {Ball}, L., \& {Skj{\ae}raasen}, O. 1999, Astroparticle Physics,
  10, 31, \dodoi{10.1016/S0927-6505(98)00041-3}

\bibitem[{Kishishita {et~al.}(2009)Kishishita, Tanaka, Uchiyama, \&
  Takahashi}]{kishishitalongterm2009}
Kishishita, T., Tanaka, T., Uchiyama, Y., \& Takahashi, T. 2009, \apj, 697, L1

\bibitem[{{Madsen} {et~al.}(2017){Madsen}, {Beardmore}, {Forster}, {Guainazzi},
  {Marshall}, {Miller}, {Page}, \& {Stuhlinger}}]{cross_calibaration}
{Madsen}, K.~K., {Beardmore}, A.~P., {Forster}, K., {et~al.} 2017, \aj, 153, 2,
  \dodoi{10.3847/1538-3881/153/1/2}

\bibitem[{Malizia {et~al.}(2021)Malizia, Fiocchi, Natalucci, Sguera, Stephen,
  Bassani, Bazzano, Ubertini, Pian, \& Bird}]{Malizia_intergral}
Malizia, A., Fiocchi, M., Natalucci, L., {et~al.} 2021, Universe, 7,
  \dodoi{10.3390/universe7050135}

\bibitem[{{Mattox} {et~al.}(1996){Mattox}, {Bertsch}, {Chiang}, {Dingus},
  {Digel}, {Esposito}, {Fierro}, {Hartman}, {Hunter}, {Kanbach}, {Kniffen},
  {Lin}, {Macomb}, {Mayer-Hasselwander}, {Michelson}, {von Montigny},
  {Mukherjee}, {Nolan}, {Ramanamurthy}, {Schneid}, {Sreekumar}, {Thompson}, \&
  {Willis}}]{Mattox1996}
{Mattox}, J.~R., {Bertsch}, D.~L., {Chiang}, J., {et~al.} 1996, \apj, 461, 396,
  \dodoi{10.1086/177068}

\bibitem[{{Paredes} {et~al.}(2000){Paredes}, {Mart{\'\i}}, {Rib{\'o}}, \&
  {Massi}}]{Paredes2000}
{Paredes}, J.~M., {Mart{\'\i}}, J., {Rib{\'o}}, M., \& {Massi}, M. 2000,
  Science, 288, 2340, \dodoi{10.1126/science.288.5475.2340}

\bibitem[{Ray {et~al.}(2011)Ray, Kerr, Parent, Abdo, Guillemot, Ransom, Rea,
  Wolff, Makeev, Roberts, Camilo, Dormody, Freire, Grove, Gwon, Harding,
  Johnston, Keith, Kramer, Michelson, Romani, Parkinson, Thompson, Weltevrede,
  Wood, \& Ziegler}]{Ray_2011}
Ray, P.~S., Kerr, M., Parent, D., {et~al.} 2011, \apjs, 194, 17,
  \dodoi{10.1088/0067-0049/194/2/17}

\bibitem[{Rea {et~al.}(2011)Rea, Torres, Caliandro, Hadasch, van~der Klis,
  Jonker, M{\'e}ndez, \& Sierpowska-Bartosik}]{readeep2011}
Rea, N., Torres, D.~F., Caliandro, G.~A., {et~al.} 2011, \mnras, 416, 1514

\bibitem[{Sarty {et~al.}(2011)Sarty, Szalai, Kiss, Matthews, Wu, Kuschnig,
  Guenther, Moffat, Rucinski, Sasselov, Weiss, Huziak, Johnston, Phillips, \&
  Ashley}]{Sarty2011}
Sarty, G.~E., Szalai, T., Kiss, L.~L., {et~al.} 2011, \mnras, 411, 1293

\bibitem[{Sironi \& Spitkovsky(2009)}]{Sironi2009}
Sironi, L., \& Spitkovsky, A. 2009, \apj, 698, 1523,
  \dodoi{10.1088/0004-637x/698/2/1523}

\bibitem[{Sironi \& Spitkovsky(2011)}]{Sironi2011}
---. 2011, \apj, 741, 39, \dodoi{10.1088/0004-637x/741/1/39}

\bibitem[{Takahashi {et~al.}(2009)Takahashi, Kishishita, Uchiyama, Tanaka,
  Yamaoka, Khangulyan, Aharonian, Bosch-Ramon, \& Hinton}]{takahashistudy2009}
Takahashi, T., Kishishita, T., Uchiyama, Y., {et~al.} 2009, \apj, 697, 592

\bibitem[{{Takata} {et~al.}(2014){Takata}, {Leung}, {Tam}, {Kong}, {Hui}, \&
  {Cheng}}]{takata2014}
{Takata}, J., {Leung}, G. C.~K., {Tam}, P.~H.~T., {et~al.} 2014, \apj, 790, 18,
  \dodoi{10.1088/0004-637X/790/1/18}

\bibitem[{{Tavani} \& {Arons}(1997)}]{1997ApJ...477..439T}
{Tavani}, M., \& {Arons}, J. 1997, \apj, 477, 439, \dodoi{10.1086/303676}

\bibitem[{{Tavani} {et~al.}(1996){Tavani}, {Hermsen}, {van Dijk}, {Strickman},
  {Zhang}, {Foster}, {Ray}, {Mattox}, {Ulmer}, {Purcell}, \&
  {Coe}}]{Tavani1996}
{Tavani}, M., {Hermsen}, W., {van Dijk}, R., {et~al.} 1996, \aaps, 120, 243.
\newblock \doarXiv{astro-ph/9611200}

\bibitem[{Torres {et~al.}(2012)Torres, Rea, Esposito, Li, Chen, \&
  Zhang}]{torresmagnetarlike2012}
Torres, D.~F., Rea, N., Esposito, P., {et~al.} 2012, \apj, 744, 106

\bibitem[{{van Dijk} {et~al.}(1996){van Dijk}, {Bennett}, {Bloemen}, {Collmar},
  {Connors}, {Diehl}, {Hermsen}, {Lichti}, {McConnell}, {Much}, {Schoenfelder},
  {Steinle}, {Strong}, \& {Tavani}}]{Dijk1996}
{van Dijk}, R., {Bennett}, K., {Bloemen}, H., {et~al.} 1996, \aap, 315, 485.
\newblock \doarXiv{astro-ph/9604136}

\bibitem[{{Volkov} {et~al.}(2021){Volkov}, {Kargaltsev}, {Younes}, {Hare}, \&
  {Pavlov}}]{Volkov2021}
{Volkov}, I., {Kargaltsev}, O., {Younes}, G., {Hare}, J., \& {Pavlov}, G. 2021,
  \apj, 915, 61, \dodoi{10.3847/1538-4357/abfe0e}

\bibitem[{{Yamaguchi} \& {Takahara}(2012)}]{Yamaguchi2012}
{Yamaguchi}, M.~S., \& {Takahara}, F. 2012, \apj, 761, 146,
  \dodoi{10.1088/0004-637X/761/2/146}

\bibitem[{Yoneda {et~al.}(2020)Yoneda, Makishima, Enoto, Khangulyan, Matsumoto,
  \& Takahashi}]{yoneda2020}
Yoneda, H., Makishima, K., Enoto, T., {et~al.} 2020, Phys. Rev. Lett., 125,
  111103, \dodoi{10.1103/PhysRevLett.125.111103}

\bibitem[{{Zdziarski} \& {Gierli{\'n}ski}(2004)}]{Zdziarski2004}
{Zdziarski}, A.~A., \& {Gierli{\'n}ski}, M. 2004, Progress of Theoretical
  Physics Supplement, 155, 99, \dodoi{10.1143/PTPS.155.99}

\end{thebibliography}

\end{document}